\documentclass[12pt,preprint]{aastex}
\slugcomment{Accepted for Publication in ApJS} %Not to appear in Nonlearned J., 45.}
\shorttitle{NEP Source Catalog}
\shortauthors{Hwang et al.}

\begin{document}

\title{An Optical Source Catalog of the North Ecliptic Pole Region\altaffilmark{1}}

\author{Narae Hwang\altaffilmark{2,11}, Myung Gyoon Lee\altaffilmark{2}, Hyung Mok Lee\altaffilmark{2},
Myungshin Im\altaffilmark{2}, Taehyun Kim\altaffilmark{2},
Hideo Matsuhara\altaffilmark{3}, Takehiko Wada\altaffilmark{3}, Shinki Oyabu\altaffilmark{3}, Soojong Pak\altaffilmark{4},
Moo-Young Chun\altaffilmark{5}, Hidenori Watarai\altaffilmark{6}, Takao Nakagawa\altaffilmark{3},
Chris Pearson\altaffilmark{3,7}, Toshinobu Takagi\altaffilmark{3}, \\Hitoshi Hanami\altaffilmark{8},
and Glenn J. White\altaffilmark{9,10}}
\altaffiltext{1}{Based on observations obtained with MegaPrime/MegaCam, a joint project of CFHT and CEA/DAPNIA, at the Canada-France-Hawaii Telescope (CFHT) which is operated by the National Research Council (NRC) of Canada, the Institut National des Science de l'Univers of the Centre National de la Recherche Scientifique (CNRS) of France, and the University of Hawaii.}
\altaffiltext{2}{Astronomy Program, Department of Physic and Astronomy, 
Seoul National University, Seoul 151-747, Korea}
\altaffiltext{3}{Institute of Space and Astronautical Science,
Japan Aerospace Exploration Agency, Sagamihara, Kanagawa 229-8510, Japan}
\altaffiltext{4}{Department of Astronomy and Space Science, Kyung Hee University,
Yongin-si, Gyeonggi-do 446-701, Korea}
\altaffiltext{5}{Korea Astronomy and Space Science Institute,
61-1, Hwaam-Dong, Yuseong-Gu, Daejeon 305-348, Korea}
\altaffiltext{6}{Office of Space Applications, Japan Aerospace Exploration Agency, Tsukuba, Ibaraki 305-8505, Japan}
\altaffiltext{7}{ISO Data Centre, ESA, Villafranca del Castillo, Madrid, Spain}
\altaffiltext{8}{Iwate University, 3-18-8 Ueda, Morioka 020-8550, Japan}
\altaffiltext{9}{Astrophysics Group, Department of Physics, The Open University, Milton Keynes, MK7 6AA, UK}
\altaffiltext{10}{Space Science \& Technology Department, CCLRC Rutherford Appleton Laboratory,
Chilton, Didcot, Oxfordshire, OX11 0QX, UK}
\altaffiltext{11}{e-mail: nhwang@astro.snu.ac.kr}

\begin{abstract}

We present a five ($u^*$,$g'$,$r'$,$i'$,$z'$) band optical photometry catalog of the sources
in the North Ecliptic Pole (NEP) region based on deep observations made with MegaCam at CFHT.
The source catalog covers about 2 square degree area centered at the NEP and
reaches depths of about 26 mag for $u^*, g', r'$ bands, about 25 mag for $i'$ band, and
about 24 mag for $z'$ band ( 4 $\sigma$ detection over an 1 \arcsec aperture).
The total number of cataloged sources brighter than $r'= 23$ mag is about 56,000
including both point sources and extended sources.
From the investigation of photometric properties using the color-magnitude diagrams and color-color
diagrams, we have found that the colors of extended sources are mostly $(u^*-r') < 3.0$ and $(g'-z') > 0.5$.
This can be used to separate the extended sources from the point sources reliably,
even for the faint source domain where typical morphological classification schemes hardly work efficiently.
We have derived an empirical color-redshift relation of the red sequence galaxies using the Sloan Digital Sky Survey data.
By applying this relation to our photometry catalog and searching for any spatial overdensities,
we have found two galaxy clusters and one nearby galaxy group.

\end{abstract}

\keywords{galaxies: general --- galaxies: photometry --- galaxies: clusters ---
catalogs}

\section{Introduction}

The North Ecliptic Pole (NEP) is an undistinguished region in the sky, located at
$\alpha = 18^h 00^m 00^s$, $\delta = +66\arcdeg 33\arcmin 38\arcsec$.
It is, however, a very special region since many astronomical satellites
have accumulated a large number of exposures over this location since
Earth-orbiting satellites must point their fixed solar panels to the Sun and
be in their orbits over the ecliptic poles.
Unlike the South Ecliptic Pole (SEP) region where the South Atlantic Anomaly and the LMC
prevent the clear view of the extragalactic sky at certain wavelengths,
the NEP region suffers very little or no obscuration by foreground Galactic sources.
The galactic coordinates of the NEP are $l \approx 96.4\arcdeg$, $b \approx +29.8\arcdeg$ and
the foreground extinction in this direction is $E(B-V) = 0.047$ \citep{sch98}.
Therefore,
the NEP is a good target for deep, unbiased, contiguous surveys for extragalactic objects such as galaxies, galaxy clusters, and AGNs.

The $ROSAT$ All-Sky Survey (RASS) is one of the most representative surveys of the NEP region \citep{vog99}.
The $ROSAT$ was the first X-ray imaging satellite to survey the entire sky.
A source catalog of X-ray sources that
were extracted from a large area surrounding the NEP was constructed from this RASS data \citep{hen01,hen06}.
From the followup observations and optical counterpart investigations,
\citet{gio03} reported that most X-ray sources in the NEP region are AGNs ($\sim 49\%$), stars ($\sim 34\%$),
and galaxy groups/clusters ($\sim 14\%$).
\citet{mul01} also found a supercluster of galaxies in the NEP region using the RASS data and
suggested that some galaxy clusters in this region are part of the supercluster at z = 0.087.
The NEP region was also observed in the 1.5 GHz band by \citet{kol94}.
Using this radio source catalog, \cite{brk99} investigated the correlation between
the radio sources and the RASS X-ray sources, and
identified optical counterparts of the radio/X-ray sources.
They found that a significant number of radio loud sources are also bright in the X-ray band,
and that X-ray selection is an effective way for the search of galaxy clusters and groups.

While the X-ray luminosity is an efficient measure of the hot ionized gas
in the galaxy clusters, the star formation activities and the resultant stellar mass
in galaxies can be estimated from the optical and infrared flux.
Recently, a new infrared space telescope, named AKARI, was launched in February 2006.
AKARI is expected to give exceedingly higher quality data than the previous infrared
space missions such as {\it Infrared Astronomical Satellite} \citep{neu84} and
{\it Infrared Space Observatory} \citep{kes96}.
The AKARI data will also complement the successful {\it Spitzer Space Telescope} \citep{wer04} data
by providing better wavelength coverage over the $8 - 24 \mu$m band, which is not available from the {\it Spitzer}.
As one of its major science programs, AKARI is currently carrying out deep near-to-mid infrared surveys
over the wide area around the NEP through two major survey programs, `NEP-Deep' and `NEP-Wide.'
The key scientific objectives of these NEP surveys are to unveil the dusty star formation history
of the Universe, the mass assembly and large scale structure evolution,
and the nature of the cosmic infrared background (CIRB).
For further information on these NEP Programs, please refer to \citet{mat06}.

Combined with these space-borne infrared surveys, the optical survey of the NEP region constitutes
the multifrequency dataset that is essential for the studies of cosmic star formation history.
Many researchers have investigated optical counterparts of radio and X-ray sources in this region,
but, for the imaging data, they mostly relied on the COSMOS scan (e.g., Brinkmann et al. 1999) or Digitized Sky Survey images
of the second Palomar Observatory Sky Survey (POSS-II) plates,
only complemented by some independent but patchy observations (e.g., Gioia et al. 2003).
Although the POSS-II data are excellent in complete coverage of the whole sky,
the usefulness of the data are still limited by the low resolution with $\sim 1.0 \arcsec$ per pixel scale
and the shallow depth with the limiting magnitude of $r_{\rm F} \sim 21.0$ mag \citep{gal04}.
In this study, therefore, we provide the first optical catalog of sources in the NEP region
based on deep observations in $u^*, g', r', i', z'$ bands obtained with MegaCam at CFHT.
This optical catalog will be an important part of multifrequency datasets of galactic and extragalactic
sources in the NEP region.
It will also serve as a basis of the AKARI NEP survey mission by providing the information
on the optical counterparts of the infrared sources.
This paper is composed as follows: We describe our observations in Section 2 and
data reduction procedures in Section 3.
We present a bright source catalog in Section 4.1, photometric properties of the sources in the NEP region
in Section 4.2, and galaxy number counts in Section 4.3.
The result of galaxy cluster search using the color of red sequence galaxies is given in Sections 4.4 and 4.5.
Some photometric properties of other X-ray sources are presented in Section 4.6 and
our primary results are summarized in the last section.

\section{Observation}\label{obs}

The observation of the North Ecliptic Pole (NEP) region was carried out over nine photometric nights
between August 22 and September 22, 2004 using the 3.6m CFHT telescope located at Mauna Kea, Hawaii.
We used a wide field imager MegaCam at the telescope primary focus MegaPrime.
The MegaCam is composed of 36 2048 $\times$ 4612 CCD's
covering about 1$\arcdeg \times 1\arcdeg$ area with 0.185 arc seconds resolution per pixel \citep{bou03}.
The $u^*,g',r',i',z'$ filter system provided with MegaCam is basically the same as that used by 
the Sloan Digital Sky Survey (SDSS) \citep{yor00} except for the $u^*$ filter
which is designed to take advantage of the improved UV extinction condition at the CFHT site.
Therefore, the photometric data generated from this observation are presented 
in the CFHT unique $u^*$ and the SDSS $g',r',i',z'$ system.
For further information on these filter systems used with MegaCam, see the `Filter set' section in http://www.cfht.hawaii.edu/Instruments/Imaging/MegaPrime/specsinformation.html page.

\begin{figure}
\plotone{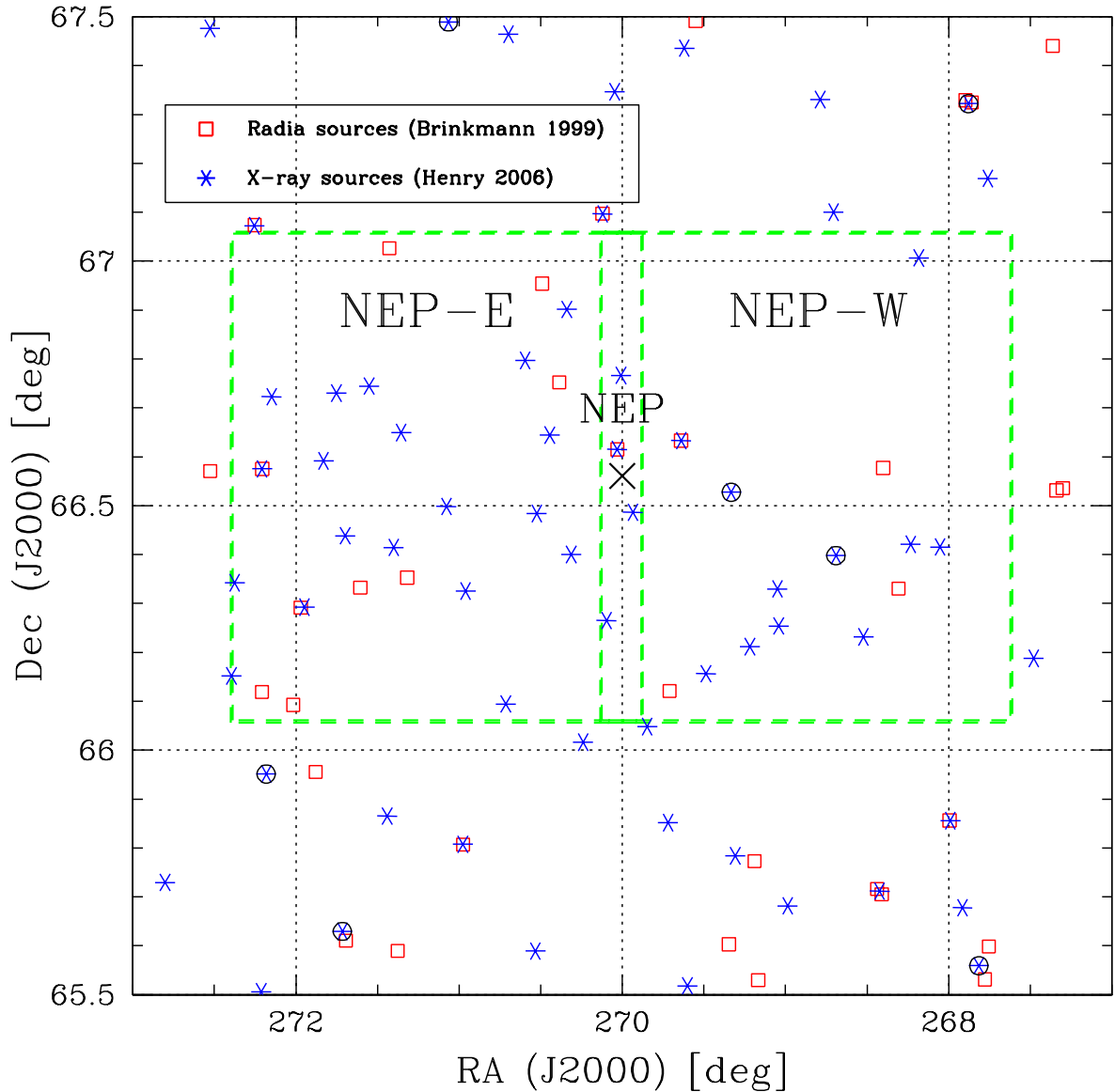}
\caption{The observation field map around the North Ecliptic Pole (NEP) (dashed line boxes).
Squares and asterisks represent, respectively, the known radio \citep{brk99} and X-ray \citep{hen06} sources.
Circles with an asterisk represent galaxy groups or clusters classified by \citet{gio03}
and cataloged in \citet{hen06}. The large cross at the center shows the location of NEP.
There are about 40 radio and/or X-ray sources found in our two fields of observation, NEP-E (East) and
NEP-W (West).}
\label{map}
\end{figure}

As shown in Figure \ref{map}, the observed fields are composed of two fields separated
by about 1$\arcdeg$ away from each other in the East-West direction, covering about two square degrees in total.
The NEP-E (East) field was observed with five filters, $u^*,g',r',i',z'$ 
and the other field, NEP-W (West), was observed with only four filters, $g',r',i',z'$.
Each field was covered twice with a 15$\arcsec$ offset for dithering in each filter with total exposure times of
3600 sec for $u^*$ (1800 $\times$ 2), 2400 sec for $g'$ (1200 $\times$ 2), 3000 sec for $r'$ (1500 $\times$ 2),
3000 sec for $i'$ (1500 $\times$ 2), and 3600 sec for $z'$ (1800 $\times$ 2).
The raw images were processed using Elixir system by CFHT staff.
The Elixir is a collection of programs, databases, and other tools specialized in processing and evaluation of the
large mosaic data \citep{mag04}.
The overall quality of the preprocessed images is good with a typical seeing of $0.7 \sim 1.1 \arcsec$.
Detailed information on the observation is listed in Table \ref{obslog}.

\section{Data Reduction}

\subsection{Pre-Photometry Processing}

For the reliable source detection and photometry, it is essential to prepare a master image
that satisfies the following two factors: (1) to obtain the required photometric depth
and (2) to give the complete areal coverage.
We used a software package SWarp\footnote{See http://terapix.iap.fr/rubrique.php?id\_rubrique=49 for further information}
written by E. Bertin at Terapix to transform each Multi Extension Fits (MEF) file
containing 36 individual CCD frames into a single fits image data.
SWarp also corrects different distortion effects between input data through the resampling process,
and then combines the resampled images to create a deep output image after flux scaling and background subtraction.

To make a master image, we combined the $g',r',i',z'$ four band data
using SWarp for the NEP-E and the NEP-W field respectively.
During the SWarp run, BILINEAR resampling method was used since
this method was found to be effective in suppressing the discontinuity effects
around the chip boundaries, while other methods such as LANCZOS3 produced more noticeable discontinuities.
We also used weighted images during SWarp run to enhance the quality of the output image.
In the weight maps we assigned zero weights to pixels with negative values.
We excluded the $u^*$ band data from the master image construction since we do not have the $u^*$ band data for the NEP-W
and the S/N is relatively lower compared to other bands.
Finally, we created two master detection images with high S/N's,
one for the NEP-E and the other for the NEP-W region.

Two MEF images for each band were processed using SWarp with the same parameters to
generate a final photometry image for the corresponding filter.
These $u^*,g',r',i',z'$ band images were made to have the same dimension and coordinates
with the master detection image of the corresponding field.
The depth of photometry attained by combining two raw images was calculated as
$4 \sigma$ flux over a circular aperture with 1 $\arcsec$ diameter.
The measured limiting magnitudes are $u^* \sim 26.0$, $g' \sim 26.1$, $r' \sim 25.6$, $i' \sim 24.7$,
and $z' \sim 23.7$ mag, as listed in Table \ref{obslog}.
After all these processings, two master detection images and nine photometry images
were prepared for the source detection and photometry.
All the information on the reduced photometry images will be available from an web site
http://astro.snu.ac.kr/\~~nhwang/index.files/nep.html.

\subsection{Source Detection \& Photometry}

We have used SExtractor \citep{ber96} to detect sources from the master detection images of the NEP-E and NEP-W fields.
A source is confirmed if it has more than five contiguous pixels above four times the background sky fluctuation.
The signals from each source were measured in the $u^*,g',r',i',z'$ band photometry images over the isophotal area previously defined
during the source detection. The photometry was made using SExtractor in dual mode operation.
This scheme enables the detection and photometry of any source
that is registered at least once in any of the $g',r',i',z'$ band images.
The total number of sources detected in our two observed NEP fields is about 130,000.

The instrumental magnitudes were transformed into standard magnitudes using the transformation information
provided and recorded in the header of the images by the CFHT staff.
During this transformation, only the calibrated magnitudes of sources that have available color information required
for the calibration are calculated and kept in the catalog.
Otherwise, we assigned dummy values (99.000) to the magnitudes of sources in the final source catalog.

\section{Results}

\begin{figure}
\plotone{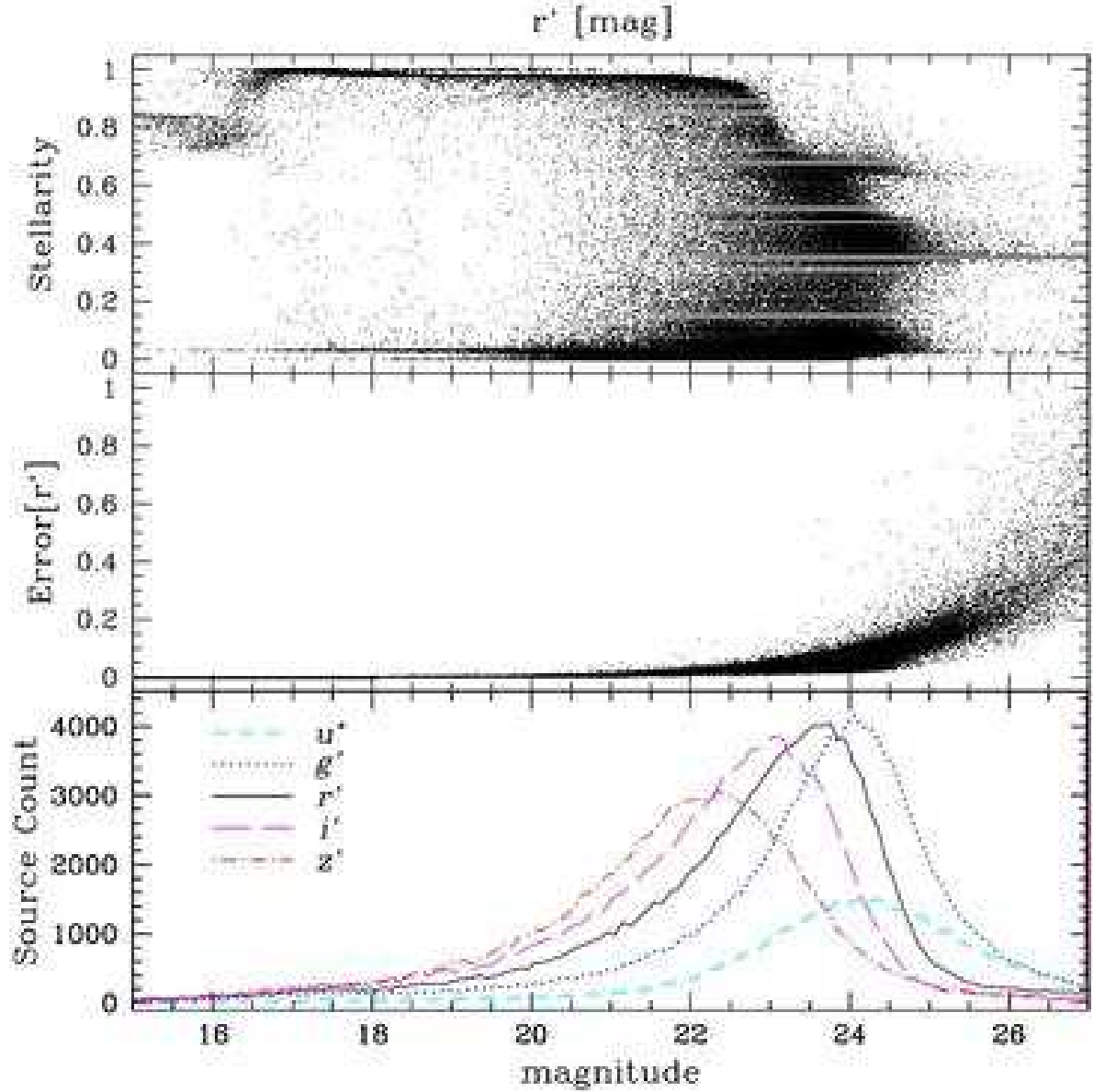}
\caption{The distribution of the stellarity index (upper panel), the $r'$ band magnitude error (mid panel),
and the source number counts in $u^*,g',r',i',z'$ bands (lower panel). The number of detected sources reaches its maximum
at about 24.3 mag for $u^*$, 24.0 mag for $g'$, 23.7 mag for $r'$, 23.0 mag for $i'$, and 22.5 mag for $z'$ band.}
\label{mhist}
\end{figure}

Figure \ref{mhist} displays a source count histogram in the $u^*,g',r',i',z'$ band along with the $r'$ band error and stellarity distributions.
This shows that the number of sources in our data increases up to $u^* \sim 24.3$, $g' \sim 24.0$, $r' \sim 23.5$,
$i' \sim 23.0$, and $z' \sim 22.2$ mag before
the incompleteness effect starts to take place and the number of the detected sources starts to decrease.
The $r'$ band magnitude error is estimated to be about 0.1 mag or less at $r' \sim 23.5$ mag
where the source count reaches its maximum.
The stellarity distribution shows how efficiently the star/galaxy classifier works in our dataset.
From this distribution, it is clearly seen that the stellarity index, which is calculated by SExtractor
based on the isophotal areas, peak intensity, and seeing information, separates the point sources (stellarity $\sim$ 1)
and the extended sources (stellarity $\sim 0$) with high confidence for the sources with $r' < 22$ mag.
One more point to be noted in this stellarity distribution is that sources with $r' < 16.5$ mag and stellarity $> 0.7$
are the results of the image saturation.
Thus the stellarity index is most relable in the magnitude range of $17 < r' < 22$ mag.

\subsection{Source Catalog}

Considering the source count and the stellarity distribution, we have decided the magnitude range of the most reliable sources
for the final bright source catalog entry, which is $r' \leq$ 23 mag.
The number of sources compiled in the final bright source catalog is about 56,000.
Table \ref{nepcat} lists a sample of the bright source catalog for reader's guide and
the full source catalog will be available electronically from the online Journal
and from an web site http://astro.snu.ac.kr/\~~nhwang/index.files/nep.html.
Followings are the short descriptions of data columns in the catalog.

Column 1 is the identification number of an optical source.

Column 2 \& 3 list, respectively, the J2000.0 right ascension (RA) and declination (DEC) of a source in degrees.
The uncertainties of the astrometric solution derived by Elixir are about $\pm 0.5 \arcsec$.

Column 4 $\sim$ 13 are the AB magnitudes and magnitude errors of a source in the $u^*,g',r',i',z'$ bands.
These are given by MAG\_AUTO and MAGERR\_AUTO parameters of SExtractor,
which are Kron-like elliptical aperture magnitudes and their errors.

Column 14 is the extraction flag of a source given as a sum of powers of 2 by SExtractor.
If a source has neighbors (flag = 1) and is blended with another source (flag = 2) and
some pixels are saturated (flag = 4),
then the final extraction flag given to the source is 7 (= 1 + 4 + 2).
Generally, a source with the extraction flag 0 gives the most reliable photometry.
For detailed information on this flag, see
the SExtractor User's Manual.

Column 15 is the stellarity index of a source calculated from a $g',r',i',z'$ combined detection image.
This index has a value between 1 (point sources) and
0 (extended sources). See Figure \ref{mhist} for the distribution of this index.

Column 16 is the ellipticity of a source calculated by SExtractor using the second order image moments.

Column 17 is the FWHM in arcseconds of a source calculated under the assumption that the source has a Gaussian profile.

Column 18 is the effective radius or a half-light radius of a source in arcseconds.
This value is computed by setting the input parameter PHOT\_FLUXFRAC = 0.5
and is given by the output parameter FLUX\_RADIUS.

Column 19 is the semi-major axis of a source in arcseconds.

Column 20 is the detected isophotal area of a source. This parameter may be used as a measure of the object's size
for the case of extended sources.

Column 21 is the field number that the photometric data of a source comes from.
The field number 1 represents the NEP-E field and 2 represents the NEP-W field.

\subsection{Color-Magnitude and Color-Color Diagrams}

We have investigated the photometric properties of the NEP sources using several color-magnitude and color-color
diagrams.
In each diagram, we use the stellarity index given by SExtractor to distinguish between point sources and extended sources:
stellarity $> 0.95$ for point sources (mostly stars) and stellarity $< 0.2$ for extended sources (mostly galaxies).
They are found to be statistically very useful tools to separate stars and galaxies
brighter than $r = 23$ mag from all kinds of sources in the catalog.

\begin{figure}
\plotone{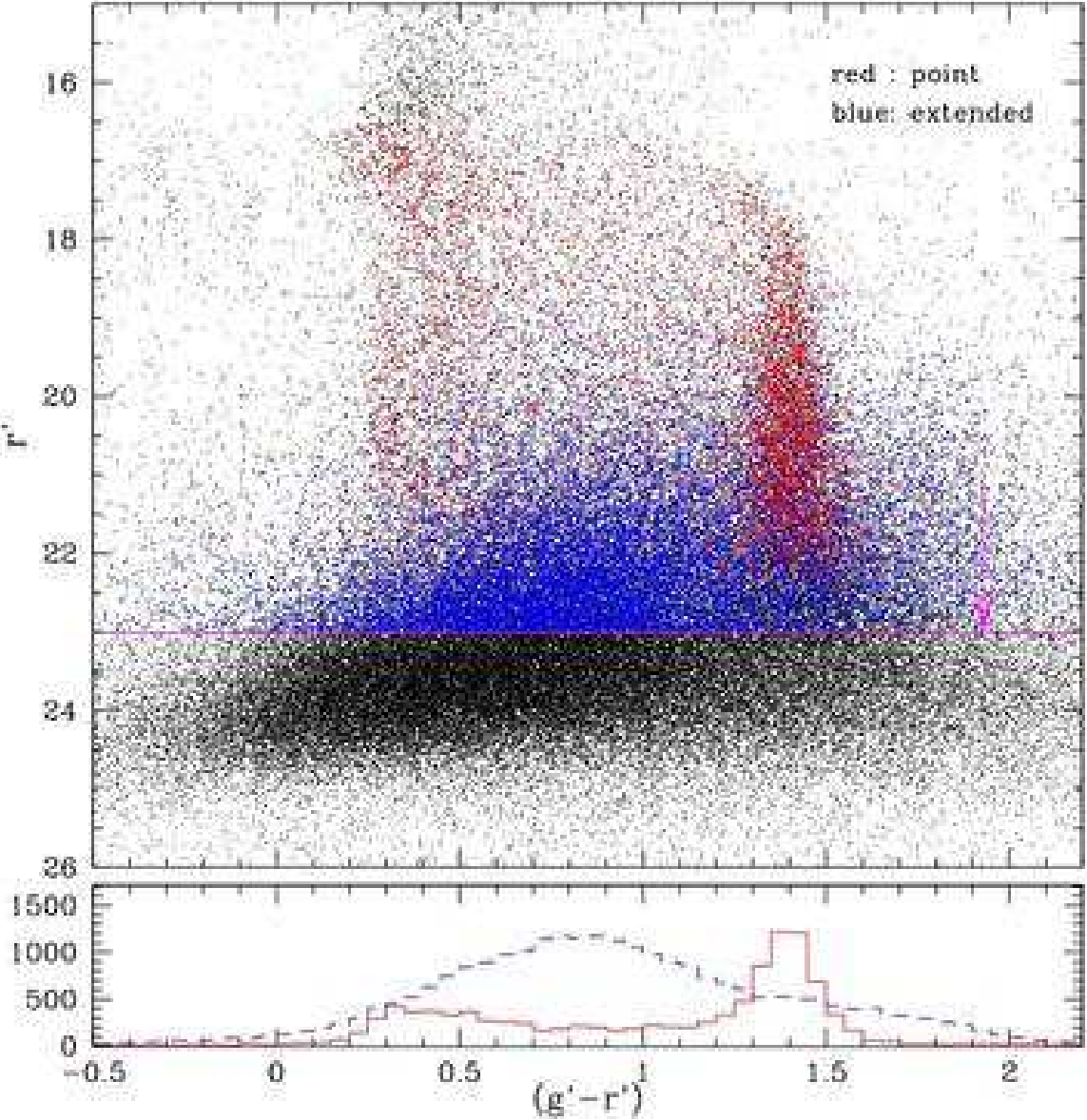}
\caption{Upper panel: $(g'-r')$ vs $r'$ color magnitude diagram (CMD) of NEP sources.
Sources selected with $r' \leq 23$ mag and $(g'-r')$ color error $< 0.1$ mag criteria are plotted in different colors:
point sources with stellarity $> 0.95$ in red and extended sources with stellarity $< 0.2$ in blue.
Black dots with $r' < 17$ mag mostly represent saturated stars.
Lower panel: $(g'-r')$ color distribution of those selected sources:
point sources in solid line and extended sources in dashed line.
Point sources and extended sources show different distributions in $(g'-r')$ color:
two peaks at $(g'-r') \approx 0.3$ (G dwarf stars) and $1.4$ (M giant stars) for point sources
and a single peak at $(g'-r') \approx 0.8$ for extended sources.}
\label{grrcmd}
\end{figure}

The $r'$ vs $(g'-r')$ color-magnitude diagram (CMD) and $(g'-r')$ color histogram
in Figure \ref{grrcmd} shows two prominent vertical sequences of
point sources, i.e., stars at $(g'-r') \sim$ 0.3 (G dwarf stars) and 1.4 (M giant stars), respectively.
The extended sources, presumably galaxies, are seen to be concentrated around the peak at $(g'-r') \sim$ 0.8
that corresponds to the $(g'-r')$ color of early type galaxies \citep{fuk95}.
The extended sources start to dominate at $r' \approx$ 22 mag and fainter.
Some of these sources may be faint stars that the SExtractor failed to classify as point sources.
However, the different $(g'-r')$ color histograms of the extended and the point sources suggest that
the majority of these faint extended sources are galaxies.

\begin{figure}
\plotone{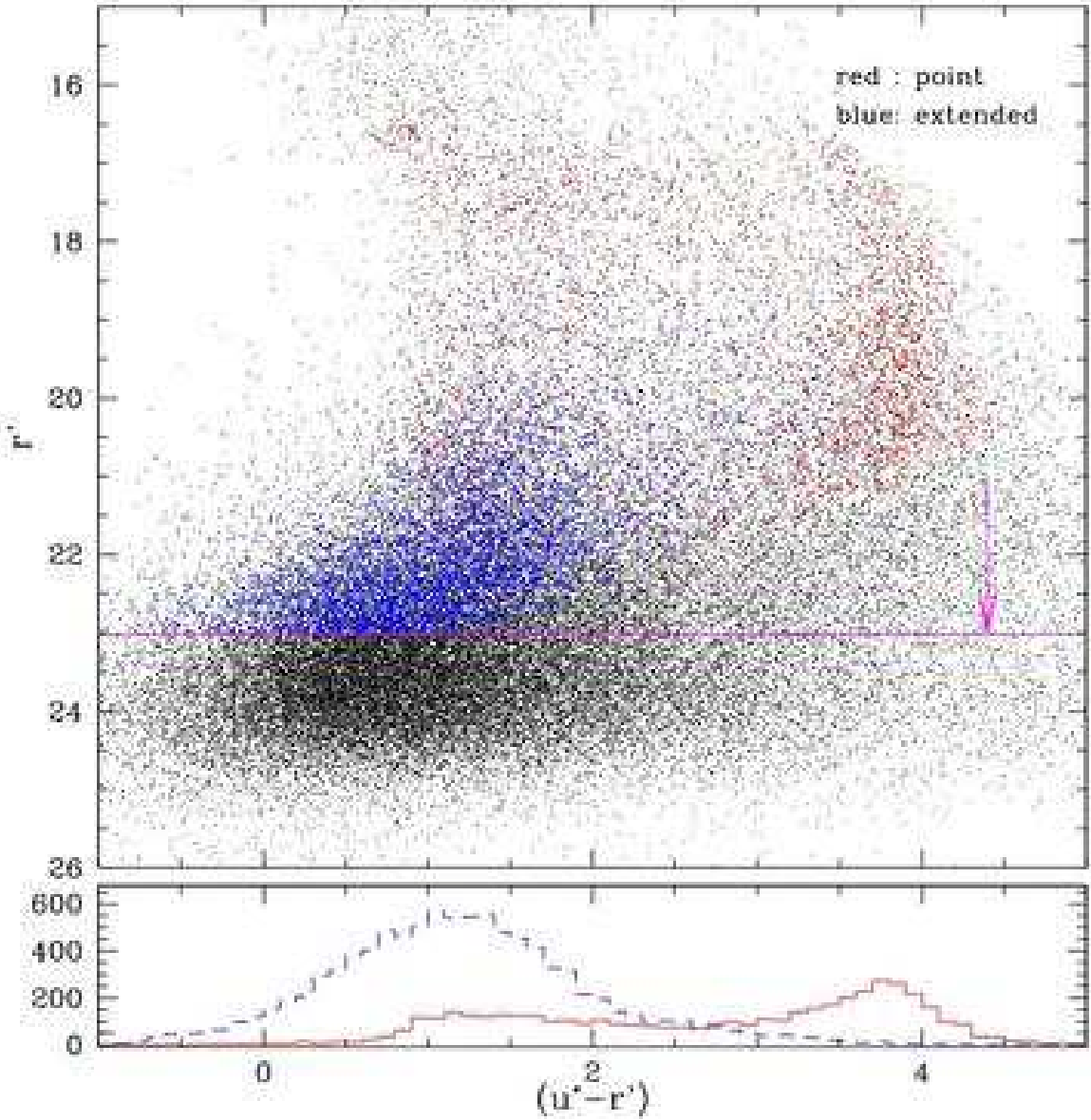}
\caption{Upper panel: $(u^*-r')$ vs $r'$ color magnitude diagram (CMD) of sources in the NEP-E field.
Sources selected with $r' \leq 23$ mag and $(u^*-r')$ color error $< 0.1$ mag criteria are plotted in different colors:
point sources with stellarity $> 0.95$ in red and extended sources with stellarity $< 0.2$ in blue.
Lower panel: $(u^*-r')$ color distribution of those selected sources:
point sources in solid line and extended sources in dashed line.}
\label{urrcmd}
\end{figure}

Figure \ref{urrcmd} displays the $r'$ vs $(u^*-r')$ CMD and $(u^*-r')$ color histogram of sources in the NEP-E field.
The CMD shows that there are many galaxies distributed around $(u^*-r') \sim$ 1.0
as shown in the $(u^*-r')$ color histogram of extended sources.
This value is consistent with the $(u^*-r')$ color of Scd type galaxies \citep{fuk95}, whereas
the $(g'-r')$ color of most galaxies in Figure \ref{grrcmd} is that of early type galaxies.
It is also noted that there is a very long redward tail in the $(u^*-r')$ color distribution of extended sources,
reaching nearly $(u^*-r') \approx$ 4 or higher.
The elliptical galaxies can be as red as $(u^*-r') \sim$ 2.8 \citep{fuk95}.
Therefore, some of those faint and red sources with $(u^*-r') > 3.0$ could be distant galaxies with redshift $z \geq 0.1$.

\begin{figure}
\plotone{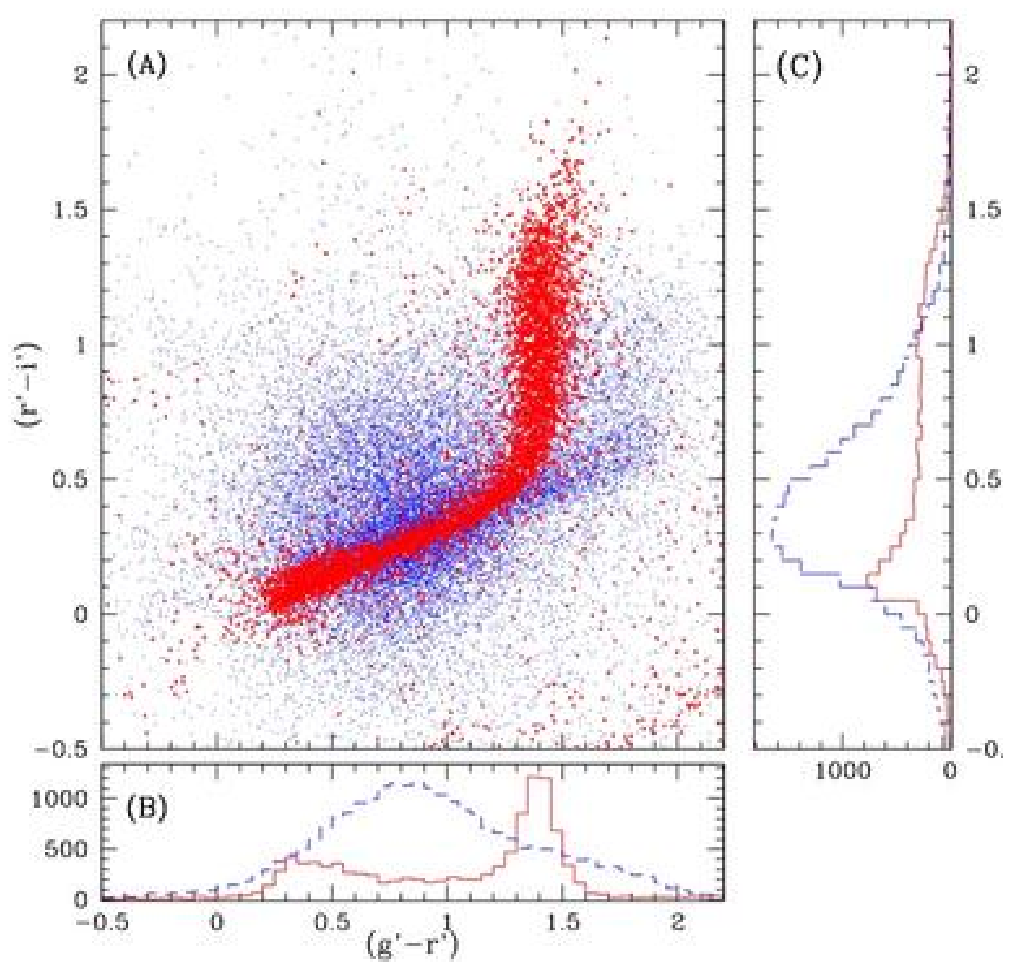}
\caption{Panel A: $(g'-r')$ vs $(r'-i')$ color-color diagram (CCD) of NEP sources.
Sources selected with $r' \leq 23$ mag, $(g'-r')$ color error $< 0.1$ mag, and $(r'-i')$ color error $< 0.1$ mag criteria
are plotted: point sources with stellarity $> 0.95$ in crosses and extended sources with stellarity $< 0.2$ in dots.
The color distribution in $(g'-r')$ color (Panel B) and $(r'-i')$ color (Panel C) of those selected sources:
point sources in solid line and extended sources in dashed line.
Please note that the $(r'-i')$ and $(g'-r')$ color sequence of point sources is very narrow and clear while
the feature of extended sources is very diffuse and broad centered at about $(g'-r') \sim 0.8$ and $(r'-i') \sim 0.3$.}
\label{grriccd}
\end{figure}

\begin{figure}
\plotone{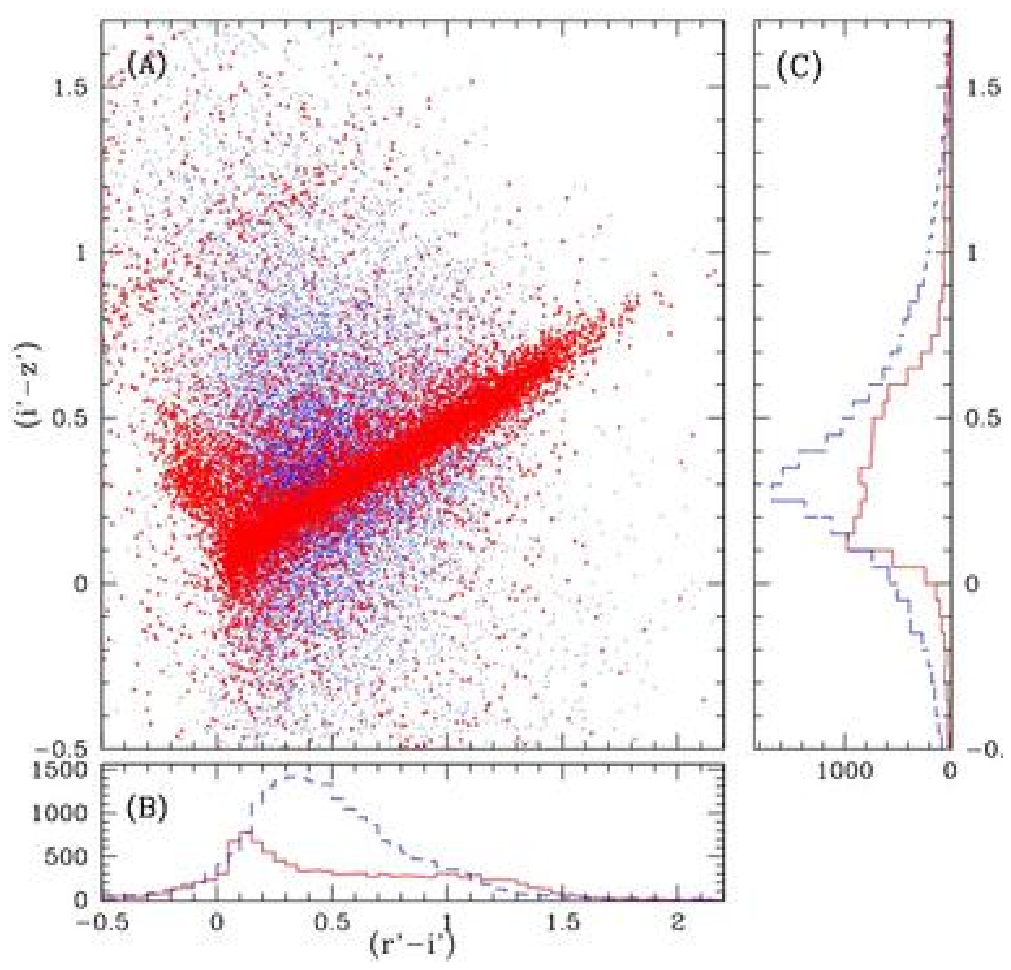}
\caption{Panel A: $(r'-i')$ vs $(i'-z')$ color-color diagram (CCD) of NEP sources.
Sources selected with $r' \leq 23$ mag, $(r'-i')$ color error $< 0.1$ mag, and $(i'-z')$ color error $< 0.1$ mag criteria
are plotted: point sources with stellarity $> 0.95$ in crosses and extended sources with stellarity $< 0.2$ in dots.
The color distribution in $(r'-i')$ color (Panel B) and $(i'-z')$ color (Panel C) of those selected sources:
point sources in solid line and extended sources in dashed line.
In this plot, point sources are distributed in a narrow straight line running from $(r'-i',i'-z') \approx (0.1,0.1)$ to $(1.8,0.8)$.}
\label{riizccd}
\end{figure}

\begin{figure}
\plotone{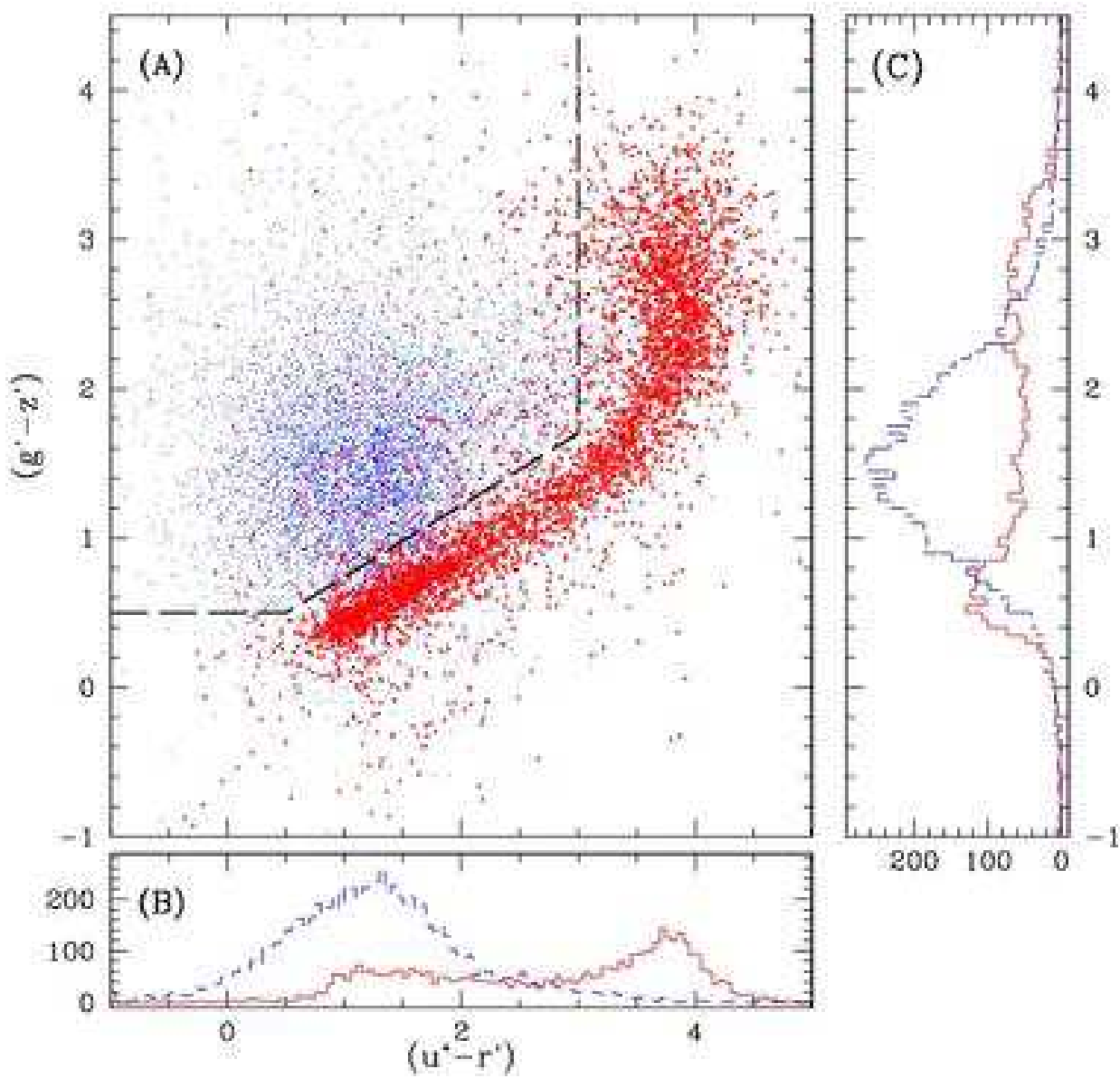}
\caption{Panel A: $(u^*-r')$ vs $(g'-z')$ color-color diagram (CCD) of NEP sources in the NEP-E field.
Selected sources with $r' \leq 23$ mag, $(u^*-r')$ color error $< 0.1$ mag, and $(g'-z')$ color error $< 0.1$ mag criteria
are plotted: point sources with stellarity $> 0.95$ in crosses and extended sources with stellarity $< 0.2$ in dots.
The dashed line in Panel A separates point sources from extended sources. See text for details.
The color distribution in $(u^*-r')$ color (Panel B) and $(g'-z')$ color (Panel C) of those selected sources:
point sources in solid line and extended sources in dashed line.}
\label{urgzccd}
\end{figure}

Figures \ref{grriccd} through \ref{urgzccd} show the characteristic distribution of point sources and extended sources
in three color-color diagrams: $(r'-i')$ vs $(g'-r')$, $(i'-z')$ vs $(r'-i')$, and $(g'-z')$ vs $(u^*-r')$.
The most prominent feature in these color-color diagrams (CCD's) is very distinguishable
sequences of point sources, i.e. stars.
In Figure \ref{grriccd}, the stellar sequence in a flipped `L' shape has a very narrow width of about 0.2 dex
in the $(r'-i')$ vs $(g'-r')$ color space.
This tight sequence is also well represented by a straight line in Figure \ref{riizccd}.
In these diagrams, extended sources are found to populate in a relatively well constrained color space
centered at a certain color.
It is still clear that we can not possibly separate extended sources from point sources
by simply constraining colors in the $(r'-i')$ vs $(g'-r')$ and the $(i'-z')$ vs $(r'-i')$ color-color space.
However, Figure \ref{urgzccd} shows that the $(g'-z')$ vs $(u^*-r')$ color-color combination
enables us to define the two exclusive spaces that are mostly populated by stars and galaxies, respectively.
The boundary of these regions can be drawn by combination of three straight lines connecting
$(u^*-r',g'-z')$ = $(-1.0,0.5)$, $(0.5,0.5)$, $(3.0,1.7)$, and $(3.0,5.0)$.
This is a very useful tool for separating faint galaxies from stars to search for distant galaxy clusters
or to compute the 2D correlation function of galaxies.

\subsection{Galaxy Number Counts}

\begin{figure}
 \plotone{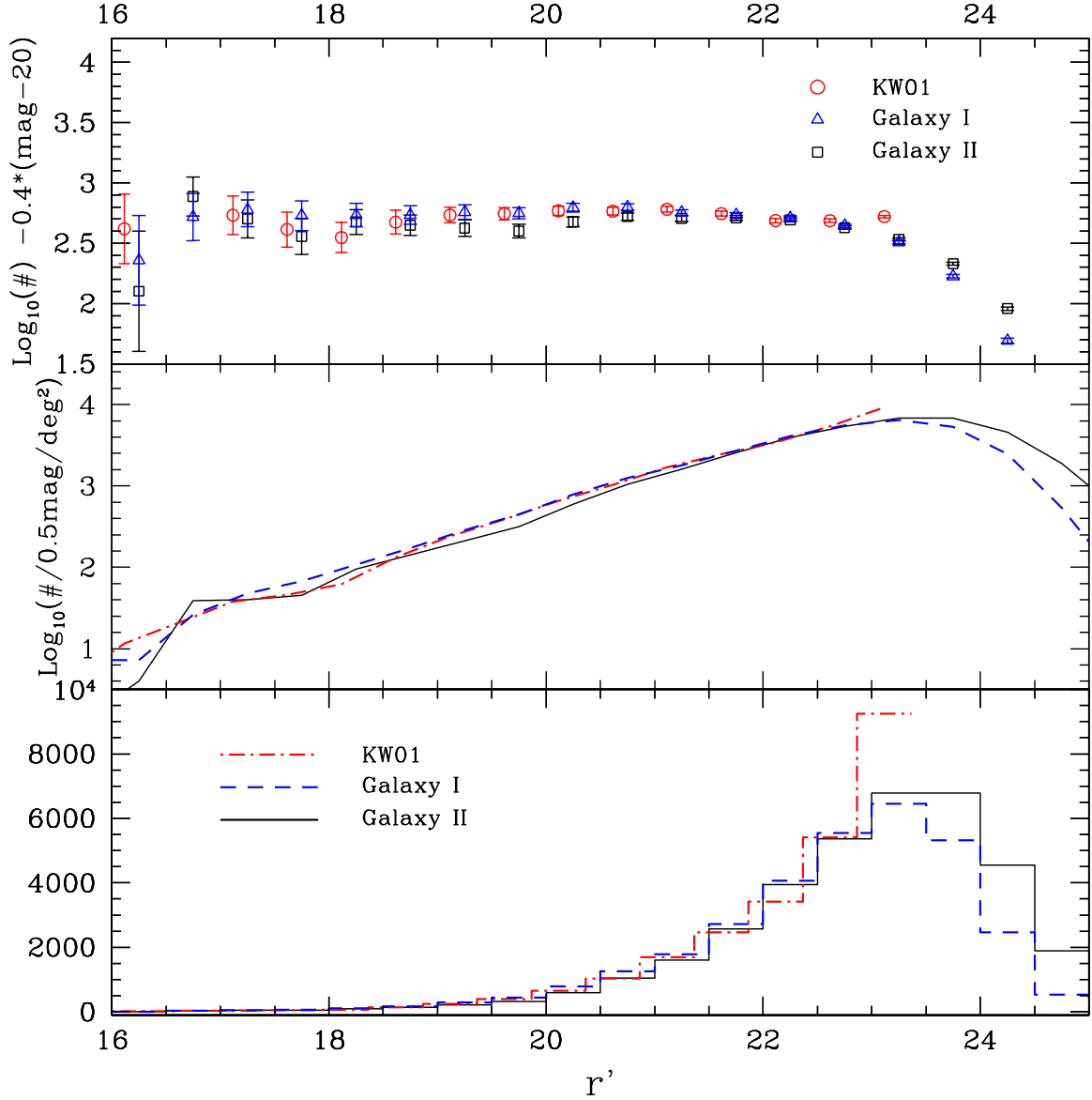}
 \caption{The number counts of galaxies in the NEP region, normalized to one square degree area.
The \citet{kum01} (KW01) data in R band are also shown in all the
three panels for comparison with our data (Galaxy I \& II)
assuming a simple relation of $r' \simeq R + 0.24$. Galaxy I is
defined to be sources with stellarity $< 0.2$ while Galaxy II is
selected based on its position on the $(u^*-r')$ vs $(g'-z')$ CCD.
It is clear that the overall shapes and
patterns of galaxy number counts are generally in good agreement
with \citet{kum01} result.}
 \label{galcount}
\end{figure}

We have investigated the galaxy number counts using the NEP optical source catalog.
To select galaxies efficiently from the photometry catalog, we adopted two different definitions of galaxies:
(1) `Galaxy I' sources with stellarity $< 0.2$ and
(2) `Galaxy II' sources that belong to the designated area where the extended sources appeared to occupy in the $(u^*-r')$ vs $(g'-z')$ CCD
as denoted by the dashed line in Figure \ref{urgzccd}.
The results of number counts for Galaxy I and Galaxy II sample, as shown in Figure \ref{galcount},
show a consistent rise from $r' = 17.5$ mag
up to about $r' = 23$ mag and then a steep down turn due to the incompleteness at $r' \geq 23$ mag.
From the lower and mid panels of Figure \ref{galcount}, it is
clear that Galaxy II sample outnumbers Galaxy I sample in $r' > 23$
mag. This difference between number counts is mostly due to the
deteriorating reliability of stellarity index in $r' \geq 23$ mag,
as shown in the upper panel of Figure \ref{mhist}. Therefore, the
Galaxy I sample is likely to lose a large number of faint galaxies
compared to the Galaxy II sample.

In Figure \ref{galcount}, we also plotted the R-band galaxy number count of \citet{kum01} for comparison
after a simple transformation into $r'$ band.
\citet{kum01} derived the number count of extended sources by modeling the number count distribution of
point-like sources (star) and then subtracting the extrapolated model count of point-like sources
from the detected source count.
On the other hand, we defined galaxies as a certain part of all the detected sources that
satisfy a given stellarity condition (Galaxy I) or a given two-color criterion (Galaxy II).
Nonetheless, it is clear from the plot that the number count by \citet{kum01} is generally consistent with
our result except for the differences in the faint magnitude domain
where the incompleteness becomes significant.

As shown in the mid panel of Figure \ref{galcount}, the logarithmic number counts defined as 
$Log(Count/0.5mag/deg^2)$ also display very similar slopes over the range of
$r' = 18 \sim 22$ mag.
Simple linear square fits over that magnitude range returned the slope d(LogN)/dm = 0.387 for Galaxy I,
0.408 for Galaxy II, and 0.397 for \citet{kum01} data.
The errors of the fitted slopes are about $\pm 0.010$ for all cases.
Therefore, the slopes of logarithmic galaxy number counts are in good agreement with that of \citet{kum01}
within the errors.

However, there are some differences in several minor features between the logarithmic profiles of galaxy number count.
This is more clearly shown in the upper panel of Figure \ref{galcount}.
It is apparent that there are two small dips in Galaxy II data:
one at $r' \approx 17.5 \sim 18$ mag and another smooth one at $r' \approx 19 \sim 20$ mag.
The first dip also appears to exist in \citet{kum01} data
but the second smooth and shallow dip could not be identified from \cite{kum01} data in this plot.
However, the magnitude of the second dip is roughly coincident with the break point at $R \approx 19 \sim 20$ mag
as shown in their Figure 4
\footnote{Figure 4 of \citet{kum01} is constructed using the galaxy number count made with 0.25 mag bin.
But the number count data in a table published in the same paper, which we used for Figure \ref{galcount},
is made with 0.5 mag bin.
The use of 0.5 mag bin instead of 0.25 mag bin in the number count is considered to cause the smoothing out the feature.}.

\subsection{Galaxy Clusters}

We have carried out a galaxy cluster search using our optical photometry catalog.
Among numerous galaxy cluster finding methods available, we adopted a simplified version of
the `C4 cluster finding algorithm' by \citet{mil05} that utilizes a seven-dimensional position and color space.
In this study, we use only a three-dimensional color space constructed based on
$(g'-r')$, $(r'-i')$, and $(i'-z')$ colors to select cluster member galaxies and then
we investigate the spatial distribution of the selected galaxies to find any overdensity of these galaxies
in small regions.
Finally, the CMDs of any overdense region are consulted to check whether the red sequence of galaxies
is apparent before we identify the region as a galaxy cluster.

This approach requires the definition of color ranges spanned by the cluster member galaxies
with various redshifts and richness classes before the actual application to the photometric data.
We have used the Sloan Digital Sky Survey (SDSS) \citep{yor00} data for this purpose.
The SDSS project provides a huge amount of multiband photometric and spectroscopic data
over a large sky area.
This enables us to select cluster member galaxies using the spectroscopy information and
to reduce the possible contamination by field galaxies and stars.
One more advantage of using the SDSS dataset to define color ranges occupied by cluster member galaxies
is that the SDSS filter system is the same as the MegaCam filter system of CFHT
except for the $u^*$ filter (see Section \ref{obs} for details).
Therefore, we have searched the SDSS database and retrieved photometric and spectroscopic data of nearby galaxy clusters
for the calibration of the red sequence colors in the SDSS filter system.

\subsubsection{Nearby Galaxy Clusters in SDSS}
\begin{figure}
\plotone{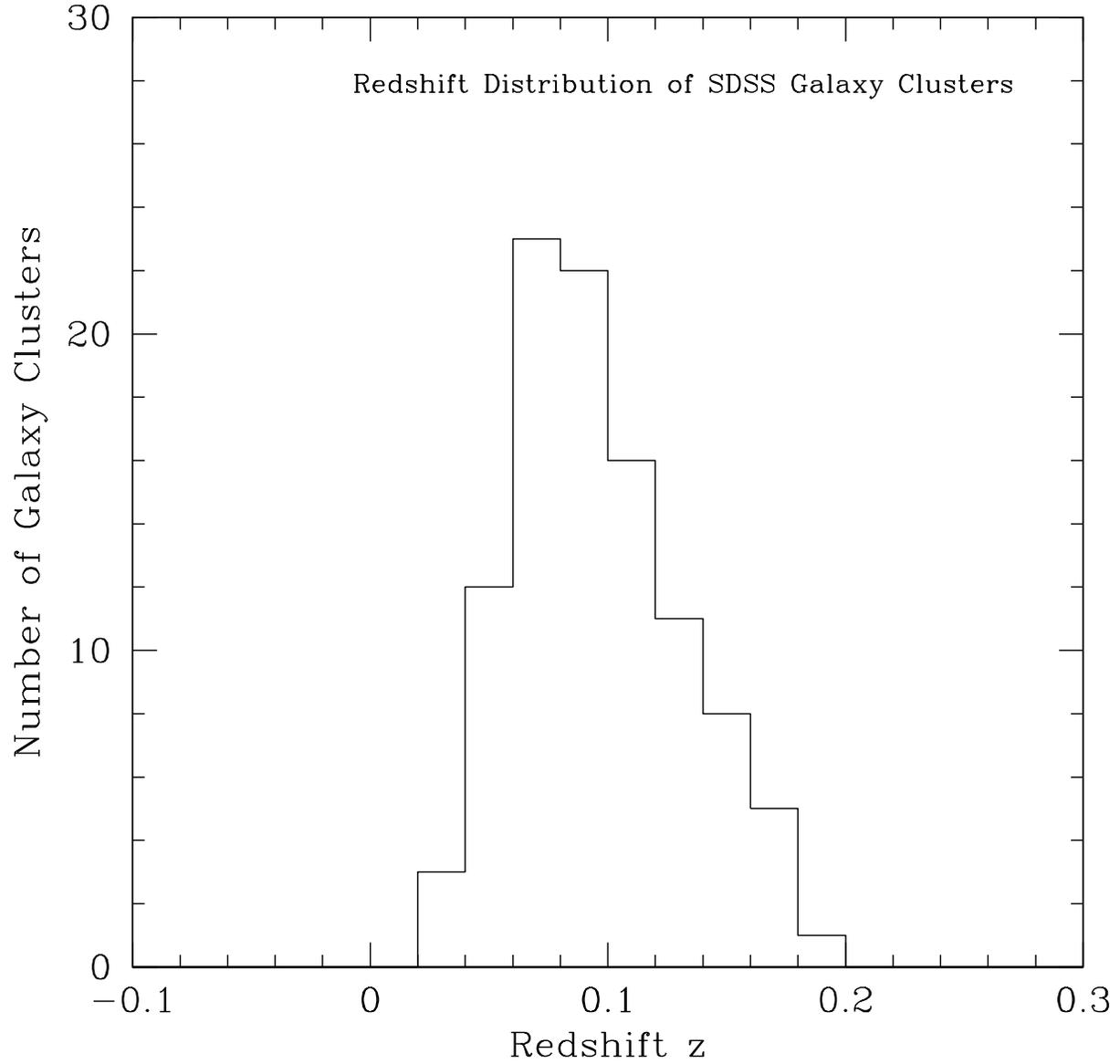}
\caption{The redshift distribution of 101 galaxy clusters whose photometric and spectroscopic data
of member galaxies were retrieved from the SDSS DR5 data archive.}
\label{sdsszdist}
\end{figure}

We have searched the SDSS DR5 database \citep{ade07} and found 101 nearby Abell galaxy clusters whose
$u',g',r',i',z'$ \footnote{Please note that the $u'$ filter is the Sloan system,
which is different from the CFHT $u^*$ system. See Section \ref{obs} for details.}
photometric and spectroscopic data of member galaxies are available.
Figure \ref{sdsszdist} displays the redshift distribution of these cluster galaxies.
The redshift of the selected galaxy clusters runs from 0 to 0.2 with a peak at $z \sim 0.07$.
After the data retrieval from the SDSS data archive,
we have selected member galaxies of each cluster based on the velocity distribution
and the spatial separation from the cluster center.
The total number of the selected member galaxies for the 101 clusters is about 5,700.
From the cluster member galaxies, only the early type galaxies were selected using the `$fracDev$'
and `$eclass$' parameters provided by the SDSS archive:
$fracDev$ $> 0.8$ and $eclass$ $<0$ (for further information on these constraints, see Bernardi et al. 2005).
Application of these parameters finally returned about 1,700 early type galaxies in 88 clusters and
the number of early type galaxies in each cluster runs from 4 (A2192) to 73 (A2199)
depending on the redshift and the richness class.

\begin{figure}
\plotone{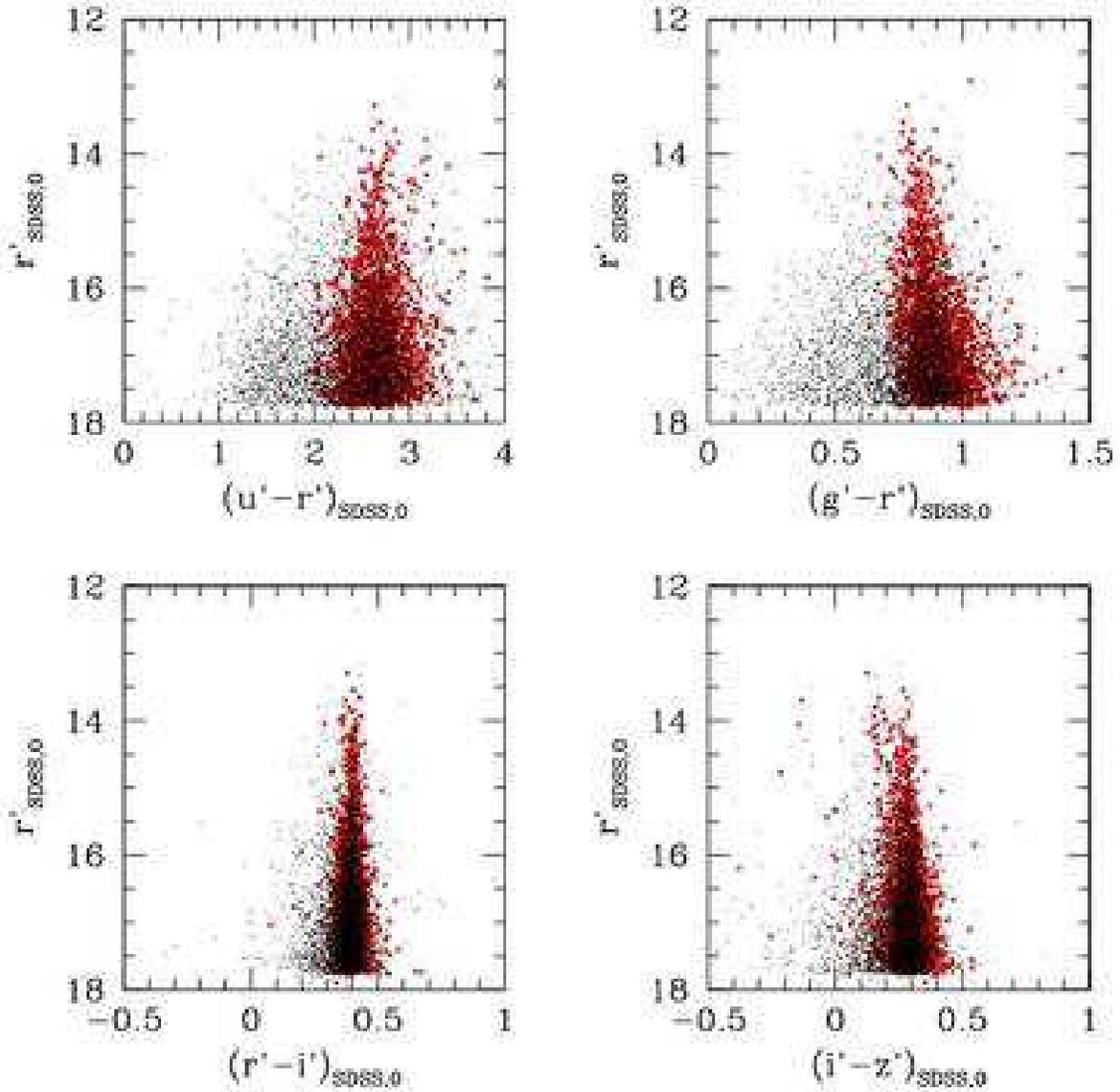}
\caption{The color-magnitude diagrams (CMD) of galaxies in 101 nearby ($z<0.2$) galaxy clusters
in the SDSS data archive.
All the photometric data used in these diagrams are foreground reddening corrected.
Early type galaxies are plotted in crosses.
Please note the relatively narrow and well-defined red sequence of early type galaxies in each CMD.}
\label{sdsscmd}
\end{figure}

\begin{figure}
 \plotone{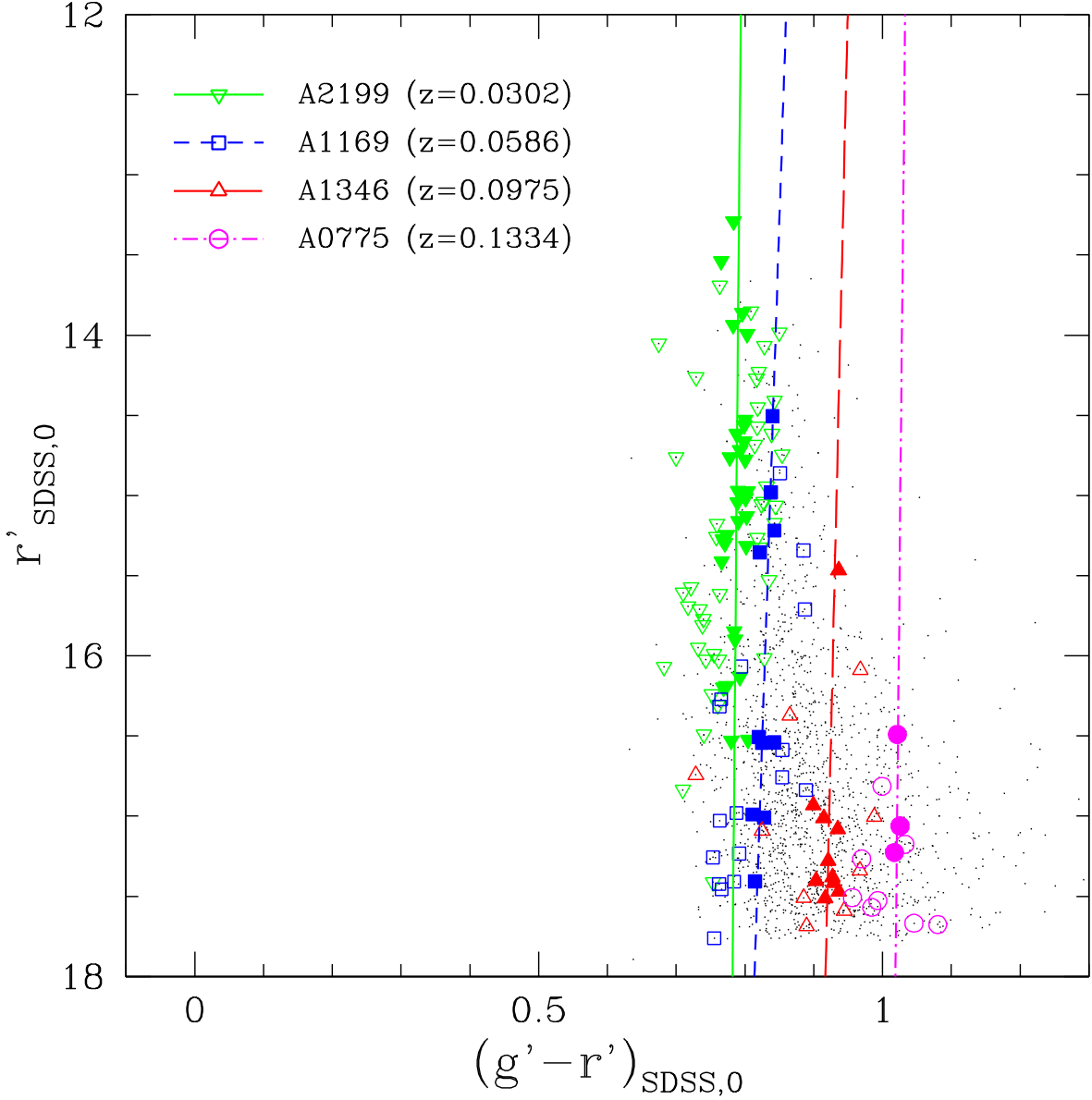}
 \caption{The representative $(g'-r')_0$ color red sequences of four galaxy clusters: A2199 (inverted triangle, $z=0.0302$),
 A1169 (square, $z=0.0586$), A1346 (triangle, $z=0.0975$), and A775 (circle, $z=0.1334$).
 The color of the red sequence was derived by fitting the colors of early type galaxies in each cluster
 marked by the filled symbols while the open symbols were rejected from the fit.
The red sequence of a galaxy cluster shifts redward as the corresponding
 redshift increases.}
 \label{sdssred}
\end{figure}

Figure \ref{sdsscmd} displays the CMDs of the cluster galaxies in
the SDSS color system. From these diagrams, it is seen that most early
type galaxies in the clusters lie within a well-defined and narrow
color range with a width of $\leq 0.2$ dex in $(g'-r')_0$, $(r'-i')_0$,
and $(i'-z')_0$. Although the $(u'-r')_0$ vs $r'_0$ CMD shows a rather large
dispersion, it still shows a strong concentration at $2.0 \leq
(u'-r')_0 \leq 3.0$. The colors of these narrow sequences of early
type galaxies have been used as indicators of the clusters'
redshifts \citep{gla00}.
We used the SDSS photometry data of member galaxies in four clusters (A2199, A1166, A1349, and A775)
to derive the relation between the clusters' redshift and the $(g'-r')_0$ color of the sequence
as shown in Figure \ref{sdssred}.
From the linear fit after repeating one sigma clipping twice, we derived a linear
and empirical relation between the redshift and the $(g'-r')_0$ color of the fitted sequences
at $r'_0 = 18$ mag (hereafter $(g'-r')_{r18,0}$) as follows:
\begin{equation}
\label{zeqn}
 {\rm Redshift}\ [z] = (0.415 \pm 0.044) \times (g'-r')_{r18,0} - (0.286 \pm 0.039)
\end{equation}
Therefore, the galaxies in nearby clusters occupy a certain space in a multi-color parameter space,
which can be used to separate galaxies in the clusters from field galaxies
and to estimate the approximate redshift of the cluster using the $(g'-r')_{r18,0}$ color of the sequence.

\subsubsection{Galaxy Cluster Search in the NEP Field}

To find galaxy clusters based on the results shown in Figure \ref{sdsscmd}
and to estimate the redshift using Equation \ref{zeqn}, the Galactic foreground reddening was corrected
for our photometry data using \citet{sch98},
assuming $A_{r'} \approx 2.751 E(B-V)$, $E(g'-r') \approx 1.042 E(B-V)$, $E(r'-i') \approx 0.665 E(B-V)$, and
$E(i'-z') \approx 0.607 E(B-V)$.
After some tests on the CFHT data,
we have defined color ranges for the cluster galaxies as follows:
$0.6 < (g'-r')_0 < 1.1$, $0.1 < (r'-i')_0 <0.4$, $0.0< (i'-z')_0 <0.4$.
We did not use $(u^*-r')_0$ color for this color space definition because of the unavailability of $u^*$ band data
for the NEP-W field and the general low S/N in the $u^*$ band in the NEP-E field.
According to a test performed after adopting the range $2.0 < (u^*-r')_0 < 3.0$ as the fourth constraining color,
the efficiency of finding the cluster galaxies turned out to be comparable to that of another test run
using the three color combination of $(g'-r')_0$, $(r'-i')_0$, and $(i'-z')_0$.
This three color combination approach also enables a homogeneous search of cluster galaxies
over our two data fields.

Possible cluster galaxies were selected by constraining the colors of galaxies
with the predefined parameters of $(g'-r')_0$, $(r'-i')_0$, and $(i'-z')_0$.
Using their RA and Dec information, we constructed spatial number density maps of the selected galaxies.
Then we identified 13 possibly overdense regions in the NEP field.
Over these 13 regions, CMDs of galaxies in $(g'-r')_0$, $(r'-i')_0$, and $(i'-z')_0$ color were constructed using the source catalog
to find any feature that resembles the red sequence of galaxies in a cluster.
From this investigation, we have identified two galaxy clusters and one nearby galaxy group,
and estimated their redshifts using the $(g'-r')_0$ colors of the red sequences.
The CMDs of those galaxy clusters and the group are presented in Figures \ref{zone1} $-$ \ref{zone6}.
More details about them are discussed below.

\subsection{Galaxy Clusters and Groups in X-ray/Radio Source Catalogs}

Some galaxy clusters or groups
were found and reported in the previous studies made
by using the X-ray and the radio band data of the NEP region.
\citet{hen95} found several X-ray-selected groups of galaxies in the NEP region based on RASS data
and this work was revised and extended further by \citet{hen06}.
\citet{brk99} also found that many X-ray sources in the NEP region have
counterparts in the radio bands that were observed with VLA.
From the X-ray and radio source catalogs provided by \citet{hen06} and \citet{brk99},
we have found a few galaxy clusters or groups in our data field.
Among these clusters and groups, we will discuss two galaxy clusters and a galaxy group
that were photometrically identified with our catalog.

\subsubsection{NEPX1/VLA 1801.5+6645}

\begin{figure}
 \plotone{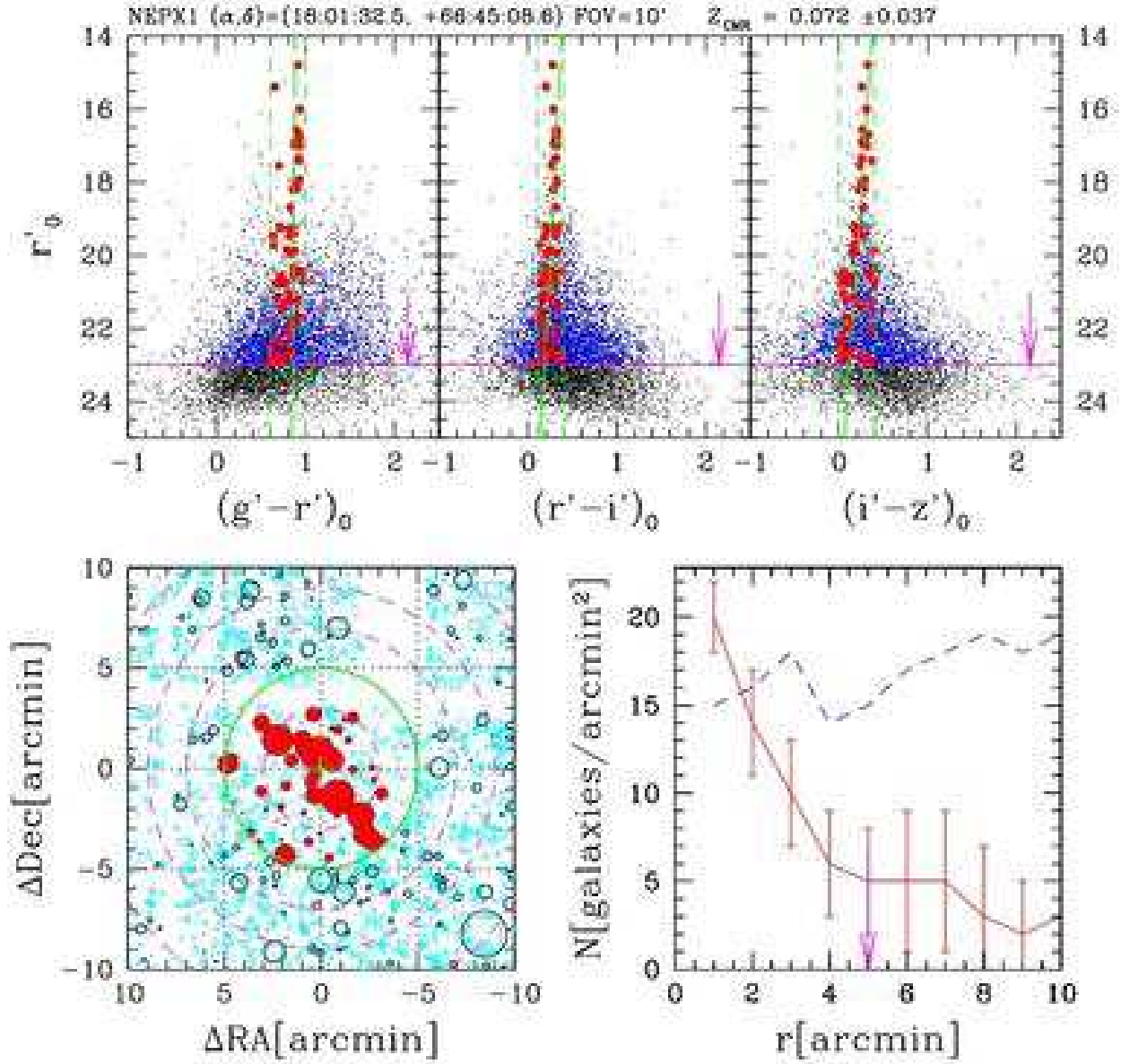}
 \caption{The CMDs (the three upper panels), the spatial distribution (lower left),
 and the number density plot (lower right) of galaxies of NEPX1.
 In each CMD, only the extended sources are plotted
 and the magnitude limit adopted for the analysis is indicated
 with an arrow.
 The color ranges used to find the cluster galaxies are shown in short dashed lines and
 the red sequence line fitted in each color is displayed in long dashed lines.
 The estimated redshift from the $(g'-r')$ color of the red sequence is indicated in the head as `$Z_{\rm CMR}$'.
 The galaxies satisfying the color cut criteria and being located within a solid circle in the lower left panel
 are represented by filled circles in the CMDs and the spatial distribution plot.
 The size of the circles in the spatial distribution plot are proportional to luminosity:
 the larger, the brighter.
 Galaxies that satisfies only the color cut criteria are plotted in open circles while other galaxies are in open squares.
 The dashed line in the density plot displays the number density profile of all galaxies around the selected area
 and the solid line shows the profile of ten times the number density of those selected galaxies and the corresponding errors.}
 \label{zone1}
\end{figure}

\begin{figure}
 \plotone{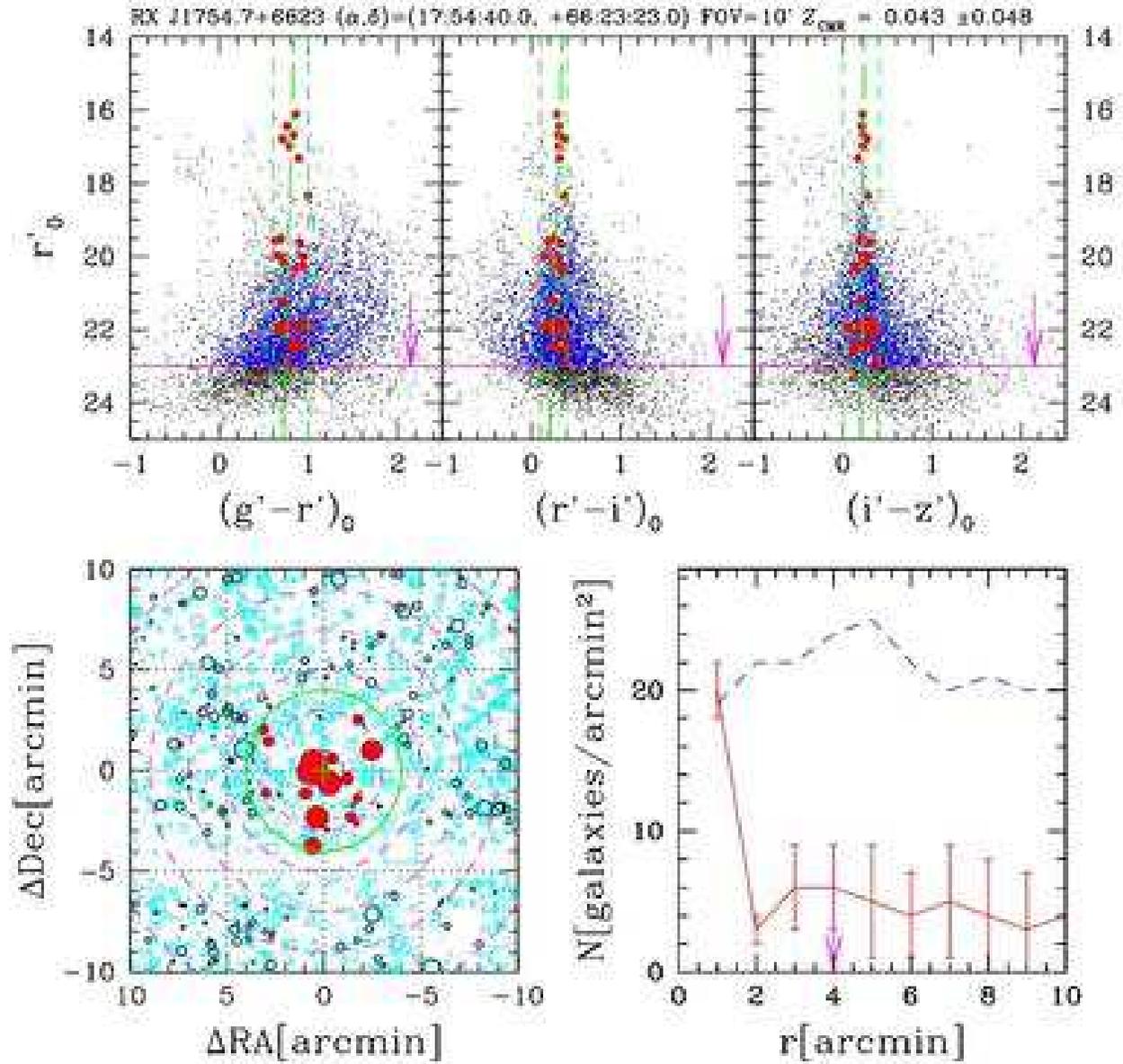}
 \caption{The CMDs (the three upper panels), the spatial distribution (lower left),
 and the number density plot (lower right) of galaxies of RX J1754.7+6623.
 There are several bright galaxies in the foreground over the many faint background galaxies
 in this region. See Figure \ref{zone1} for the legend.}
 \label{zone3}
\end{figure}

\begin{figure}
 \plotone{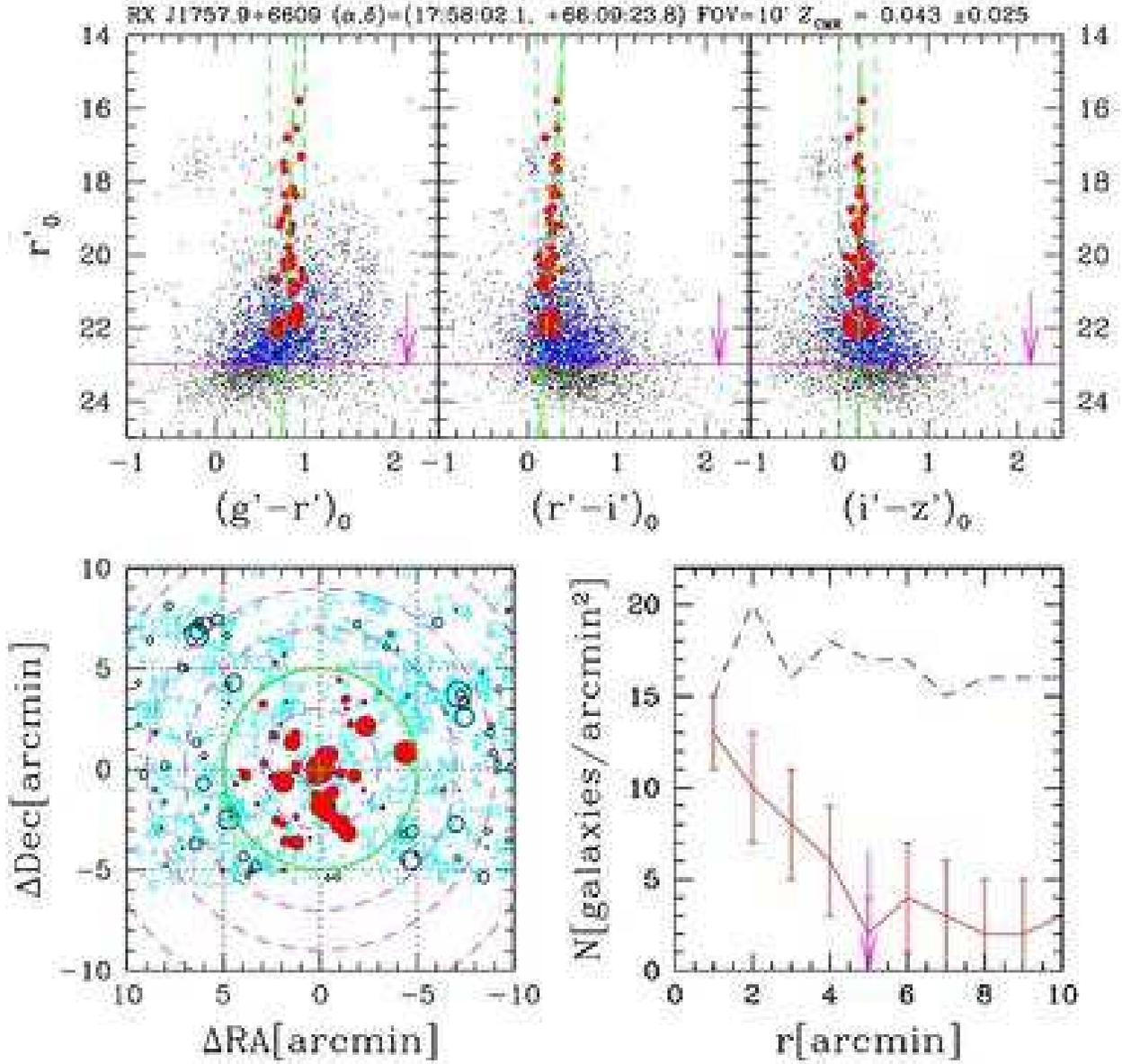}
 \caption{The CMDs (the three upper panels), the spatial distribution (lower left),
 and the number density plot (lower right) of galaxies of RX J1757.9+6609.
 See Figure \ref{zone1} for the legend.}
 \label{zone6}
\end{figure}

This is the richest galaxy cluster found in our data field,
which is easily seen in Figure \ref{zone1}.
In the upper panel of Figure \ref{zone1}, the CMDs of galaxies show very strong red sequences
running from $r'_0 \approx 16$ mag (magnitude of the third brightest galaxy) in three color domains: $(g'-r')_0$, $(r'-i')_0$, and $(i'-z')_0$.
Those galaxies belonging to the red sequences are concentrated in a very compact
region,
as shown in the lower panel of Figure \ref{zone1}.
The redshift estimated from the red sequence's $(g'-r')_{r18,0}$ color and Equation \ref{zeqn}
is about $0.072 \pm 0.037$.
This cluster was previously discovered by \citet{bur92} from ROSAT
survey data and named as NEPX1. They estimated the redshift
of the cluster to be about 0.09 from the spectroscopic observations,
which is in good agreement with our estimate.

This cluster is considered as a part of the large supercluster structure that
has been reported to exist in the NEP region \citep{has91,mul01}.
It was also detected in the radio
band from the VLA observations and was suggested as a possible
counterpart of the X-ray source RXS J180137.7+664526 by \citet{brk99}.
Although \citet{brk99} listed this source as a galaxy group,
we have reached a conclusion based on the red sequence galaxies in the CMDs
that it is a galaxy cluster rather than a galaxy group.

\subsubsection{RX J1754.7+6623}

\citet{hen06} classified this source as a galaxy cluster with redshift $z = 0.0879$.
However, the CMDs shown in Figure \ref{zone3} suggest that there
are two kinds of galaxies: (1) several bright galaxies that
form the red sequence running from $r'_0 \approx 16$ to $18$ mag
and (2) many faint galaxies in the background.
The second component of galaxies are not brighter than
$r'_0 = 19$ mag, which is about 1 mag fainter than the faintest galaxy of
the brighter component.
This indicates that those bright galaxies may happen to be located in front of the faint background galaxies
by chance.
Therefore, we suggest that these galaxies are members of a galaxy group rather than
a galaxy cluster.
The red sequence's $(g'-r')_{r18,0}$ color
is about 0.793, corresponding to an estimated redshift of $0.043 \pm 0.048$.
The relative large error may be due to the poorly determined slope of red sequence depends sensitively
on several bright galaxies with $r'_0 < 18$ mag.

\subsubsection{RX J1757.9+6609}

There is a known X-ray source RX J1757.9+6609 at the similar position.
It is listed as a type 2 AGN with redshift $z = 0.4865$ in the catalog of \citet{hen06}.
A type 2 AGN is a narrow emission line (FWHM $< 2000$ km $s^{-1}$) object,
while a type 1 AGN is a broad emission line (FWHM $\geq 2000$ km $s^{-1}$) object \citep{gio03}.
The spatial number density plot (lower panel) of the red sequence galaxies
in Figure \ref{zone6}
shows some clustering in the 2D projected space.
This point was also noted by \citet{gal03} who carried out a
cluster search using the digitized Second Palomar Observatory Sky Survey (DPOSS) data.
They detected an overdensity of galaxies at this location and listed it as a galaxy cluster candidate
under the name of NSC J175751+660924, and also estimated its redshift
to be $z \approx 0.1663$ from their photometric redshift analysis.
Our photometric data
shows well developed red sequences
running from $r'_0 \approx 17.5$ mag 
in $(g'-r')_0$, $(r'-i')_0$, and $(i'-z')_0$ colors,
confirming that it is a genuine galaxy cluster.
The red sequence's $(g'-r')_{r18,0}$ color is about 0.811 and
this leads to the estimated redshift of $z = 0.043 \pm 0.025$.
Therefore, there appears to be a galaxy cluster with redshift $z < 0.2$ and
a type 2 AGN with redshift $z = 0.4875$ in the background of the same projected area.

\subsubsection{Other Galaxy Clusters}

We have identified two galaxy clusters and one galaxy group by applying the simple color cut method
to our photometry catalog.
However, there are two more galaxy clusters that are listed in the literature in our observation field
that we failed to confirm.
One is RX J1801.7+6637 ($z=0.57$) reported by \citet{bow96} and the other is RX J1757.3+6631 ($z=0.6909$)
from the catalog of \citet{hen06}.
There appear to be some weak hints of clustering around these clusters on the images.
But our photometry is not deep enough to identify any red sequence of galaxies in the CMD of the galaxies.

\subsection{The Photometric Properties of Other X-ray Sources}

\begin{figure}
 \plotone{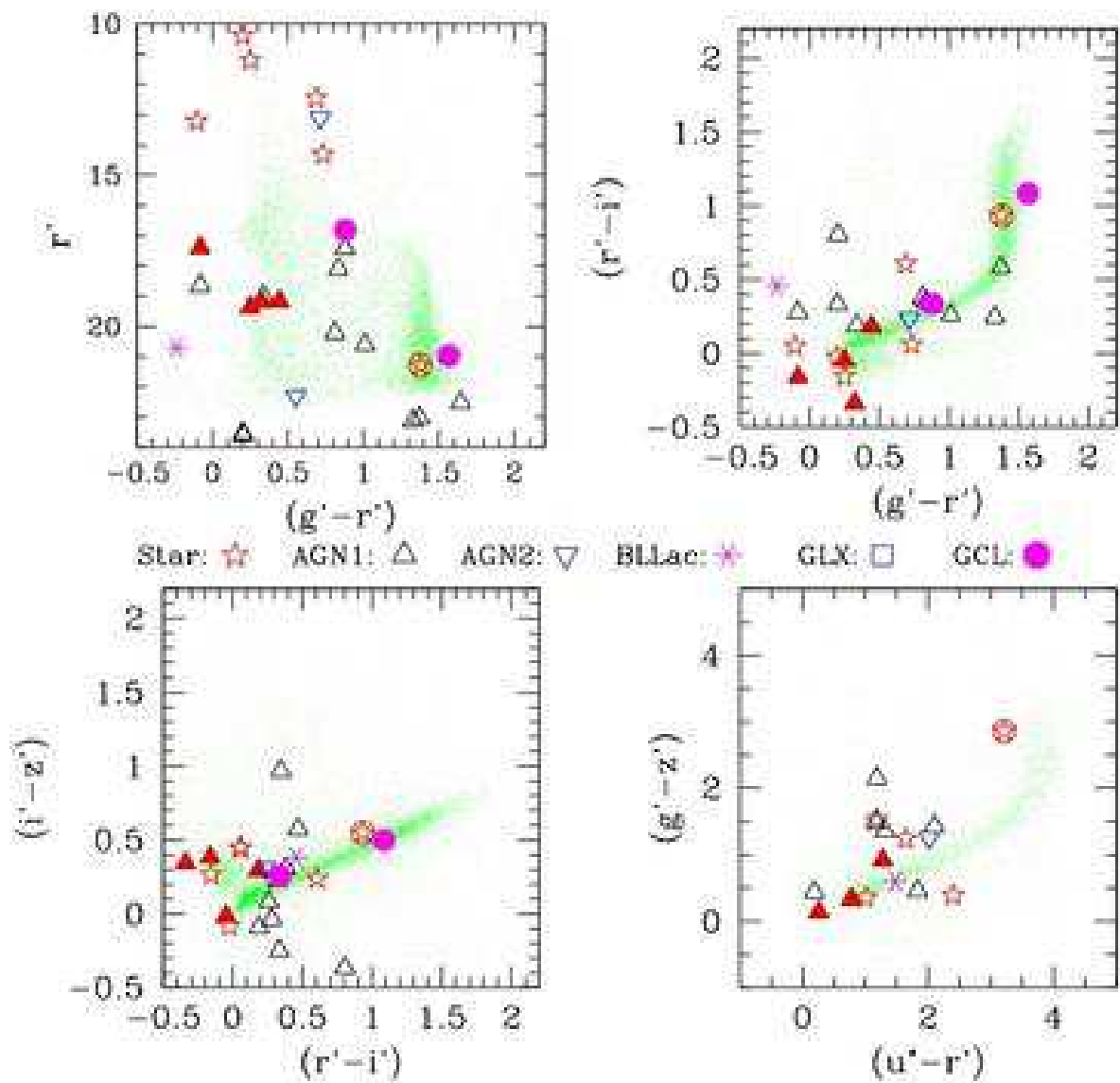}
 \caption{The photometric properties of X-ray sources. Stellar symbols for stars, triangles for type 1 AGN (AGN1),
 reversed triangles for type 2 AGN (AGN2), asterisks for BL Lac's, squares for galaxies,
 filled circles for galaxy clusters, and dots for point sources from our photometry catalog.
 The classification information of X-ray sources are from \citet{hen06}.
 The distribution of point sources is presented in each CMD or CCD for references.
 Stars marked by stellar symbols in these diagrams are suspected to suffer from saturation effect
 except for one faint star represented by a stellar symbol within a circle.
 Please note that the point-like AGN1 sources (filled triangles) are bluer than $(g'-r') \sim 0.5$.}
 \label{h06cmd}
\end{figure}

The photometric properties of various X-ray sources such as stars, AGNs, BL Lac's
could provide valuable information regarding their stellar populations and evolution.
Comparison of our photometry catalog with the X-ray source catalog of \citet{hen06}
returned about 25 possible matches.
Based on the classification information of the X-ray sources by \citet{hen06},
we have investigated the photometric property of each object class using several
CMDs and CCDs.
Figure \ref{h06cmd} shows a $(g'-r')$ vs $r'$ CMD and the three different CCDs of various optical
counterparts of X-ray sources.
For a guideline in each diagram, the distribution of point sources, which are mostly stars, is also
plotted.

In Figure \ref{h06cmd}, it is easily seen that
(1) generally, X-ray bright stars do not belong to the well defined stellar sequence in each diagram, and
(2) AGNs and BL Lac objects are not readily separated from stars or other kind of objects
in these CMDs and CCDs.
Most X-ray bright stars that are used in this comparison are very bright ($r' < 15$ mag)
and are mostly saturated except for one faint giant star as shown in the upper-left panel of Figure \ref{h06cmd}.
This may explain why these X-ray bright stars appear to be in different color domains
from other generic stars.
Even one faint star (marked by a star with a circle in Figure \ref{h06cmd}) is
bluer in $(u^*-r')$ color than normal stars.
Although AGNs and BL Lac's as a whole do not show any distinct pattern in each color,
the AGN1s with stellarity $> 0.9$, which are relatively free from wrong identification,
are found only in the blue domains of $(g'-r')$, $(r'-i')$, and $(u^*-r')$.
The $(g'-r')$ CMD also shows that any sources with $(g'-r')<0.0$ and $r' > 15 \sim 16$ mag are very likely to be AGN or BL Lac objects.

\section{Summary and Conclusion}

We have obtained $u^*,g',r',i',z'$ optical band high resolution images of
the two square degrees area centered at the NEP
with MegaCam/MegaPrime at CFHT.
From the source detection and the photometry using SExtractor,
we have compiled about 56,000 sources with $r' \leq 23$ mag,
including point and extended sources,
into the final optical photometry catalog as listed in Table \ref{cat}.
The use of the color-magnitude diagrams and color-color diagrams revealed
strikingly different photometric characteristics of
stars and galaxies.
We have found that the use of $(u^*-r')$ vs $(g'-z')$ color enables us
to clearly separate galaxies from stars and this separation does not suffer
from the uncertainties involved in the morphological classification of faint sources.
The galaxy number counts constructed from the galaxies selected based on the
$(u^*-r')$ vs $(g'-z')$ color show a nearly monotonic increase up to about $r' = 23$ mag
with a slope d(LogN)/dm $\approx 0.40 \pm 0.01$, which is in agreement with
the literature.
However, there are some changes in the slope at $r' = 17.5 \sim 18$ mag and
$r' = 19 \sim 20$ mag, which needs further studies.

Using the SDSS DR5 data of the 101 nearby Abell galaxy clusters with redshift $z < 0.2$,
we have derived a relation between the redshift and the color of the red sequences in the SDSS filter system
which is compatible with the CFHT MegaCam filter system.
Utilizing the information derived from the nearby galaxy clusters,
we have applied a simple color cut method to find galaxy clusters
in our data field, which returned two galaxy clusters and one galaxy group.
These galaxy clusters and group are also radio and X-ray sources,
which were reported by previous studies.
We have also estimated the redshift of these galaxy clusters and group using the
linear relation between the $(g'-r')_{r18,0}$ color and the redshift.
The estimated redshift is in agreement with the known value of 0.09 for galaxies in NEPX1
but it is relatively lower than the spectroscopic redshift in the literature for galaxies in RX J1754.7+6623
and the photometrically derived redshift for RX J1757.9+6609.
For RX J1754.7+6623, this underestimation of redshift may be due to the poorly determined slope of red sequence
since there are only a few bright galaxies.
For RX J1757.9+6609, the galaxy cluster seems to be overlayed on the background type 2 AGN with redshift of 0.4875.

We have compared our photometry catalog with an X-ray source catalog in the literature
to investigate the photometric properties of other X-ray sources.
The result of comparison implies that the sources with $(g'-r') < 0.0$ could be classified as
candidates of AGNs or BL Lac objects.

\acknowledgments
N.H. and T.K. are in part supported by the BK21 program of the Korean Government.
This work was in part supported by the ABRL (R14-2002-058-01000-0).

\clearpage

\begin{deluxetable}{cccccccr}
\tabletypesize{\scriptsize}
\tablecaption{Observation Log \label{obslog}}
\tablewidth{0pt}
\tablehead{
\colhead{Field} & \colhead{RA} & \colhead{DEC} & \colhead{Filter} & \colhead{Total Exp Time} & \colhead{Seeing} &
\colhead{Depth} & \colhead{Observation Date (UT)} \\
\colhead{} & \colhead{(hh:mm:ss)} & \colhead{(dd:mm:ss)} & \colhead{} & \colhead{(sec)} & \colhead{(arcsec)} &
\colhead{(AB mag)\tablenotemark{a}} & \colhead{(yyyy-mm-dd)}}
\startdata
NEP-E & 18:04:31.51 & 66:33:38.60 & $u^*$ & 1800 $\times$ 2 & 1.13 & 25.98 & 2004-09-13/14 \\
      &             &             & $g'$ & 1200 $\times$ 2 & 1.08 & 26.12 & 2004-08-22 \\
      &             &             & $r'$ & 1500 $\times$ 2 & 0.99 & 25.58 & 2004-08-23 \\
      &             &             & $i'$ & 1500 $\times$ 2 & 0.69 & 24.70 & 2004-08-23/09-12 \\
      &             &             & $z'$ & 1800 $\times$ 2 & 0.73 & 24.03 & 2004-09-07 \\
\\ \hline \\
NEP-W & 17:55:28.49 & 66:33:38.60 & $g'$ & 1200 $\times$ 2 & 1.01 & 26.12 & 2004-08-22 \\
      &             &             & $r'$ & 1500 $\times$ 2 & 0.87 & 25.58 & 2004-09-19/20 \\
      &             &             & $i'$ & 1500 $\times$ 2 & 0.99 & 24.85 & 2004-09-13/14 \\
      &             &             & $z'$ & 1800 $\times$ 2 & 0.85 & 23.71 & 2004-09-22 \\
\enddata
\tablenotetext{a}{The depth of each filter data was measured as 4 $\sigma$ flux over a circular aperture with a diameter of 1 \arcsec.}
\end{deluxetable}

\clearpage

\begin{deluxetable}{rccccccccccccccccccccccc}
\tabletypesize{\tiny}
\setlength{\tabcolsep}{0.03in}
\rotate
\tablecaption{NEP Optical Source Catalog\label{cat}\tablenotemark{a}}
\tablewidth{0pt}
\tablehead{
\colhead{ID} & \colhead{RA (J2000)} & \colhead{Dec (J2000)} & \colhead{$u^*$} & \colhead{err($u^*$)} & \colhead{$g'$} & \colhead{err($g'$)} & \colhead{$r'$} & \colhead{err($r'$)} &
\colhead{$i'$} & \colhead{err($i'$)} & \colhead{$z'$} & \colhead{err($z'$)} &
\colhead{flag} & \colhead{stellarity} & \colhead{ellipticity} & \colhead{fwhm} &
\colhead{r\_eff} & \colhead{sma} & \colhead{area} &
\colhead{field} \\ %& \colhead{remarks} \\
 & \colhead{[deg]} & \colhead{[deg]} & \colhead{[mag]} &  & \colhead{[mag]} &  & \colhead{[mag]} &  &
\colhead{[mag]} &  & \colhead{[mag]} &  &
 &  &  & \colhead{[arcsec]} &
\colhead{[arcsec]} & \colhead{[arcsec]} & \colhead{[arcsec$^2$]} &  } %& }
\startdata
      3 & 271.1438293 &  66.0613251 &  99.000 & 99.000 &  25.438 &  0.098 &  22.822 &  0.020 &  22.608 &  0.031 &  26.193 &  1.738 &     0 & 0.08 &   0.295 &   1.611 &   0.620 &   0.423 &    1.677 &  1 \\ %&  ...  \\
      5 & 271.4050293 &  66.0613251 &  99.000 & 99.000 &  24.105 &  0.033 &  22.895 &  0.020 &  22.200 &  0.019 &  99.000 & 99.000 &     0 & 0.76 &   0.042 &   0.990 &   0.460 &   0.347 &    1.814 &  1 \\ %&  ...  \\
      6 & 272.0104980 &  66.0599136 &  99.000 & 99.000 &  99.000 & 99.000 &  18.853 &  0.001 &  18.484 &  0.000 &  99.000 & 99.000 &     4 & 1.00 &   0.162 &   0.773 &   0.326 &   0.156 &    0.274 &  1 \\ %&  ...  \\
     17 & 270.6894836 &  66.0615005 &  99.000 & 99.000 &  24.819 &  0.066 &  22.994 &  0.025 &  21.750 &  0.014 &  99.000 & 99.000 &     0 & 0.28 &   0.056 &   1.373 &   0.562 &   0.389 &    1.814 &  1 \\ %&  ...  \\
     20 & 270.6847839 &  66.0614319 &  99.000 & 99.000 &  26.160 &  0.182 &  22.875 &  0.022 &  21.472 &  0.010 &  99.000 & 99.000 &     0 & 0.18 &   0.133 &   2.218 &   0.580 &   0.398 &    1.882 &  1 \\ %&  ...  \\
     24 & 271.0955811 &  66.0624542 &  23.651 &  0.040 &  23.783 &  0.026 &  22.803 &  0.019 &  22.544 &  0.028 &  99.000 & 99.000 &     0 & 0.76 &   0.110 &   1.117 &   0.531 &   0.334 &    1.437 &  1 \\ %&  ...  \\
     25 & 270.9777527 &  66.0611649 &  99.000 & 99.000 &  20.694 &  0.002 &  19.145 &  0.001 &  18.382 &  0.001 &  99.000 & 99.000 &     0 & 0.99 &   0.132 &   1.025 &   0.407 &   0.481 &    6.640 &  1 \\ %&  ...  \\
     26 & 270.6968384 &  66.0618744 &  23.781 &  0.050 &  25.055 &  0.061 &  22.911 &  0.018 &  22.476 &  0.022 &  99.000 & 99.000 &     0 & 0.88 &   0.129 &   1.175 &   0.484 &   0.318 &    1.369 &  1 \\ %&  ...  \\
     35 & 271.8403320 &  66.0610962 &  24.433 &  0.072 &  24.490 &  0.041 &  22.855 &  0.018 &  22.677 &  0.028 &  99.000 & 99.000 &     0 & 0.88 &   0.160 &   1.032 &   0.468 &   0.333 &    1.506 &  1 \\ %&  ...  \\
     37 & 270.7434998 &  66.0618744 &  99.000 & 99.000 &  24.161 &  0.036 &  22.042 &  0.011 &  20.852 &  0.006 &  99.000 & 99.000 &     0 & 0.98 &   0.090 &   0.995 &   0.506 &   0.408 &    2.772 &  1 \\ %&  ...  \\
\tablenotetext{a}{The complete version of this table is in the electronic edition of
the Journal. The printed edition contains only a sample.}
\enddata
\label{nepcat}
\end{deluxetable}

\clearpage
\thispagestyle{empty}
\setlength{\voffset}{-25mm}

\clearpage
\setlength{\voffset}{0mm}

\end{document}